\documentclass[%
reprint,
superscriptaddress,
amsmath,
amssymb,
aps,
prl,
longbibliography
]{revtex4-2}

\usepackage[T1]{fontenc}
\usepackage[utf8]{inputenc}
\usepackage{pdfpages} 
\usepackage{tikz}

\usepackage{times} 
\usepackage[unicode=true,breaklinks=true,colorlinks=true]{hyperref}
\hypersetup{citecolor=blue,urlcolor=blue}
\usepackage{graphicx}
\usepackage{bm}
\usepackage{bbold}

\usepackage{mathtools}
\usepackage{tensor}

\makeatletter 
\AtBeginDocument{\let\LS@rot\@undefined} 
\makeatother 

\begin{document}
\title{Dynamic framework for criticality-enhanced quantum sensing}
\author{Yaoming Chu}
\affiliation{MOE Key Laboratory of Fundamental Physical Quantities Measurements, Hubei Key Laboratory of Gravitation and Quantum Physics, School of Physics, Huazhong University of Science and Technology, Wuhan 430074, China}
\affiliation{International Joint Laboratory on Quantum Sensing and Quantum Metrology, Institute for Quantum Science and Engineering, Wuhan National Laboratory for Optoelectronics, Huazhong University of Science and Technology, Wuhan 430074, China}
\author{Shaoliang Zhang}
\email{shaoliang@hust.edu.cn}
\affiliation{MOE Key Laboratory of Fundamental Physical Quantities Measurements, Hubei Key Laboratory of Gravitation and Quantum Physics, School of Physics, Huazhong University of Science and Technology, Wuhan 430074, China}
\affiliation{International Joint Laboratory on Quantum Sensing and Quantum Metrology, Institute for Quantum Science and Engineering, Wuhan National Laboratory for Optoelectronics, Huazhong University of Science and Technology, Wuhan 430074, China}
\author{Baiyi Yu}
\affiliation{MOE Key Laboratory of Fundamental Physical Quantities Measurements, Hubei Key Laboratory of Gravitation and Quantum Physics, School of Physics, Huazhong University of Science and Technology, Wuhan 430074, China}
\affiliation{International Joint Laboratory on Quantum Sensing and Quantum Metrology, Institute for Quantum Science and Engineering, Wuhan National Laboratory for Optoelectronics, Huazhong University of Science and Technology, Wuhan 430074, China}
\author{Jianming Cai}
\email{jianmingcai@hust.edu.cn}
\affiliation{MOE Key Laboratory of Fundamental Physical Quantities Measurements, Hubei Key Laboratory of Gravitation and Quantum Physics, School of Physics, Huazhong University of Science and Technology, Wuhan 430074, China}
\affiliation{International Joint Laboratory on Quantum Sensing and Quantum Metrology, Institute for Quantum Science and Engineering, Wuhan National Laboratory for Optoelectronics, Huazhong University of Science and Technology, Wuhan 430074, China}
\affiliation{State Key Laboratory of Precision Spectroscopy, East China Normal University, Shanghai, 200062, China}

\date{\today}

\begin{abstract}
Quantum criticality, as a fascinating quantum phenomenon, may provide significant advantages for quantum sensing. Here we propose a dynamic framework for quantum sensing with a family of Hamiltonians that undergo quantum phase transitions (QPT). By giving the formalism of the quantum Fisher information (QFI) for quantum sensing based on critical quantum dynamics, we demonstrate its divergent feature when approaching the critical point. We illustrate the basic principle and the details of experimental implementation using quantum Rabi model. The framework is applicable to a variety of examples and does not rely on the stringent requirement for particular state preparation or adiabatic evolution. It is expected to provide a route towards  the implementation of criticality-enhanced quantum sensing. 
\end{abstract}
\maketitle
{\it Introduction.---} Quantum sensing can achieve ultra-precise measurements by exploiting the distinct features of quantum phenomena  \cite{Giovannetti2004,Giovannetti2011,Degen2017}. Apart from quantum superposition and entanglement, quantum criticality also represents a powerful resource \cite{Sachdev2011,Macieszczak2016,Rams2018,Frerot2018,Zanardi2008} for sensing, following the intuition that quantum systems around the critical points are likely to be very sensitive to minute change of physical parameters. Not only the ground state near QPT \cite{Zanardi2006,Zanardi2008,You2007,Invernizzi2008,Schwandt2009,Albuquerque2010,Wang2014,Salvatori2014,Bina2016,Rossini2018,Mirkhalaf2020,Gong2008,Gu2008,Greschner2013} but also general non-equilibrium steady-state around a dissipative phase transition \cite{Lorenzo2017,Wald2020} demonstrate divergent susceptibility to the parameters in the Hamiltonian. The dynamics of Loschmidt echo \cite{Quan2006,Mukherjee2012,Tsang2013,Hwang2019} and certain order parameters \cite{Yang2019a} associated with the ground state can also exhibit criticality enhanced parameter dependence. These critical behaviors raise intensive interest in designing quantum sensing protocols that may be powered by quantum criticality \cite{Goltsman2001,Raghunandan2018,Wang2018,Yang2019,Dutta2019,Ivanov2020,Garbe2020}.
However, the advantages provided by quantum criticality are usually hindered by the time required to prepare the ground state close to the critical points, e.g. via adiabatic evolution \cite{Rams2018}. Such a requirement imposes great challenges on the implementation of criticality-enhanced quantum sensing. A framework for quantum sensing enhanced by critical quantum dynamics that can relax the stringent requirement for state preparation would pave the way towards experimental realization of criticality-enhanced quantum sensing, which however remains largely unexplored \cite{Macieszczak2016,Skotiniotis2015,Fiderer2018}. 
In this work, we fill in this gap and provide a framework of quantum sensing based on critical quantum dynamics for a family of parameter ($\lambda$)-dependent Hamiltonians $\hat{H}_\lambda=\hat{H}_0+\lambda \hat{H}_1$. We provide a general formalism of the QFI for the associated quantum dynamics, and demonstrate that a divergent scaling in the QFI can be achieved for general initial states. Note that the QFI is equivalent to the inverse variance of the measurement, thus critical quantum dynamics enables prominent enhancement in the achievable measurement precision. Taking quantum Rabi model (QRM) as an illustrative example, we elaborate the experimentally feasible schemes that can reach the sensitivity close to the quantum Cram\'er-Rao bound given by the QFI.  Furthermore, we provide two other experimentally feasible examples including the optical parametric oscillator and the Lipkin-Meshkov-Glick (LMG) model. The framework provides a general and feasible way of exploiting quantum criticality for superior quantum sensing. 
{\it The QFI of critical quantum dynamics.---} Generally, the performance of quantum sensing associated with a parametric dynamic evolution $U_\lambda=\exp(-i \hat{H}_\lambda t)$ is determined by the optimal ability of resolving neighboring states $|\Psi_\lambda\rangle=U_\lambda|\Psi\rangle$ and $|\Psi_{\lambda+\delta\lambda}\rangle=U_{\lambda+\delta\lambda}|\Psi\rangle$ in the parameter space, which is characterized by the QFI. 
%These two states are related by $|\Psi_{\lambda+\delta\lambda}\rangle=\exp(-i J_\lambda \delta\lambda)|
%\Psi_\lambda\rangle$, where $J_\lambda=-i U_\lambda \partial_\lambda U_\lambda^\dagger$ is the local %generator of $U_\lambda$ \cite{Pang2017}. 
The QFI with respect to the parameter $\lambda$ can be formulated as $I_\lambda=4\text{Var}[h_\lambda]_{|\Psi\rangle}$, where $\text{Var}[\cdot]_{|\Psi\rangle}$ represents the variance corresponding to the initial state $|\Psi\rangle$ and $h_\lambda=U_\lambda^\dagger J_\lambda U_\lambda=iU_\lambda^\dagger\partial_\lambda U_\lambda$ is the transformed local generator with $J_\lambda=-i U_\lambda \partial_\lambda U_\lambda^\dagger$ \cite{Pang2017}.

We consider a class of Hamiltonians $\hat{H}_\lambda$, an eigenvalue equation of which can be written in the following form as \cite{Pang2014}
\begin{equation}
[\hat{H}_\lambda,\hat{\Lambda}]=\sqrt{\Delta}\hat{\Lambda}, \label{eq:general-condition}
\end{equation}
where $\hat{\Lambda}=i\sqrt{\Delta}\hat{C}-\hat{D}$ with $\hat{C}=-i [\hat{H}_0,\hat{H}_1]$, $\hat{D}=-[\hat{H}_{\lambda},[\hat{H}_0,\hat{H}_1]]$ and $\Delta$ is dependent on the parameter $\lambda$. The above relation in Eq.\eqref{eq:general-condition} usually leads to an equally spaced energy spectrum of $\hat{H}_{\lambda}$ with the energy gap $\epsilon \sim \sqrt{\Delta}$ when $\Delta>0$ \cite{supplement}. On the other hand, if $\Delta<0$, $\sqrt{\Delta}$ becomes imaginary. The non-analytical behaviors would appear at the critical point $\lambda=\lambda_c$ defined by $\Delta=0$,  implying the emergence of QPTs. Based on Eq.\eqref{eq:general-condition}, we obtain the transformed local generator as
\begin{equation}
\label{eq:generator}
h_\lambda=\hat{H}_1 t+\frac{\cos(\sqrt{\Delta}t)-1}{\Delta} \hat{C}-\frac{\sin(\sqrt{\Delta}t)-\sqrt{\Delta}t}{\Delta\sqrt{\Delta}} \hat{D}.
\end{equation}
It can be seen that $h_\lambda$ becomes divergent as $\Delta\to 0$ if $\sqrt{\Delta}t\simeq \mathcal{O}(1)$, which represents a signature of critical quantum dynamics. 
%The critical feature of $h_\lambda$ will be able to enhance the distinguishability of $|\Psi_\lambda\rangle$ and $|%\Psi_{\lambda+\delta\lambda}\rangle$, and thus yields a superior resolution for $\lambda$. 
From the transformed local generator in Eq.\eqref{eq:generator}, we obtain the QFI as follows
\begin{equation}
  \label{eq:QFI}
  I_\lambda(t)\simeq 4\frac{[\sin(\sqrt{\Delta}t)-\sqrt{\Delta}t]^2}{\Delta^3}\text{Var}[\hat{D}]_{|\Psi\rangle}. 
\end{equation}
Under the condition $\sqrt{\Delta}t\simeq \mathcal{O}(1)$, the QFI scales with $\Delta $ as $I_\lambda\sim \Delta^{-3}\sim \Delta^{-2}t^2$ and is divergent at $\Delta=0$. Such a scaling of the QFI arises from the parametric unitary evolution and holds for general initial states ${|\Psi\rangle}$ provided that $\text{Var}[\hat{D}]_{|\Psi\rangle}\simeq \mathcal{O}(1)$ or even more general mixed states \cite{supplement}, which avoids the requirement for ground state preparation. We remark that it is applicable for any Hamiltonian that satisfies the relation in Eq.\eqref{eq:general-condition}, the phase transition of which occurs across the whole spectrum. 

%In the following, we first consider the example of quantum Rabi model, and then demonstrate the general %feasibility by extending to other possible models. 
%

%
{\it Critical quantum sensing with QRM.---} We start by considering the quantum Rabi model as an illustrative example, which represents an important model in quantum optics. The model consists of a two-level system (qubit) and a bosonic field mode, which exhibits a normal-to-superradiant phase transition \cite{Hwang2015,Felicetti2020}. The system's Hamiltonian is
\begin{equation}
\label{eq:H-QRM}
\mathcal{H}_{\text{Rabi}}=\omega a^\dagger a+\frac{\omega_0}{2}\sigma_z^{(q)}-\lambda (a+a^\dagger)\sigma_x^{(q)}.
\end{equation}
Here, $\sigma_{x,z}^{(q)}$ are Pauli operators of the qubit with the transition frequency $\omega_0$ and $a$ ($a^\dagger$) is the annihilation (creation) operator of the bosonic field with the frequency $\omega$. 
The simplicity of such a quantum system together with the experimental capability of precise coherent quantum control \cite{Puebla2017} facilitates the engineering of a critical quantum sensor based on QRM. We apply a Schrieffer-Wolff transformation and 
%to the Hamiltonian $\mathcal{H}_{\text{Rabi}}\rightarrow e^{-\mathcal{S}} \mathcal{H}_{\text{Rabi}} %e^{\mathcal{S}}$ with $\mathcal{S}=(i g/2)\eta^{-\frac{1}{2}}(a+a^\dagger)\sigma_y+\mathcal{O}(\eta^{-\frac{3}
%{2}})$. In the limit of $\eta=\omega_0/\omega\to \infty$, we 
obtain an effective low-energy Hamiltonian for the normal phase $ \mathcal{H}_{np}^{({\downarrow})}=
%\langle\downarrow|e^{-\mathcal{S}} \mathcal{H}_{\text{Rabi}} e^{\mathcal{S}}|\downarrow\rangle_q
\omega [a^\dagger a-g^2(a+a^\dagger)^2/4]-\omega_0/2$ with $g=2\lambda/\sqrt{\omega\omega_0}<1$  in the limit of $\eta=\omega_0/\omega\to \infty$ \cite{supplement}. The normal-to-superradiant phase transition occurs at the critical point $g=1$. The Hamiltonian $ \mathcal{H}_{np}^{({\downarrow})}$  satisfies the relation in Eq.\eqref{eq:general-condition}, and we define ${\Delta_g}=4(1-g^2)$ which is consistent with the energy gap $\epsilon\propto |1-g^2|^{\nu z}$
with the critical exponents satisfying $\nu z=1/2$  \cite{Hwang2015}. As an example, given the two-level system in the spin-down state $|{\downarrow}\rangle_q$, we obtain the QFI for the measurement of the parameter $g$ with similar analysis by which we derive Eq.\eqref{eq:QFI} as
\begin{equation}
  \label{eq:QFI-QRM}
  I_g (t)\simeq 16g^2\frac{[\sin(\sqrt{\Delta_g}\omega t)-\sqrt{\Delta_g}\omega t]^2}{\Delta_g^3} \text{Var}[\hat{P}^2]_{|\varphi\rangle_b},
\end{equation}
where $\hat{P}=i(a^\dagger-a)/\sqrt{2}$ and $|\varphi\rangle_b$ is the initial state of the bosonic field mode \cite{supplement}. It can be seen that $I_g \to \infty$ when $g\to 1$ (i.e. $\Delta_g \to 0$), thus it would allow us to estimate the parameter with a precision enhanced by critical quantum dynamics. 
%We note that such a critical behavior of the QFI can be robust to small coherent deviations from the QRM %Hamiltonian in Eq.\eqref{eq:H-QRM} \cite{supplement}. 
The ultimate precision of quantum parameter estimation is given by quantum Cram\'er-Rao bound (namely the inverse of the QFI) requiring the optimal measurement. We hereby provide two measurement protocols that are experimentally feasible to achieve the precision of the same order as quantum Cram\'er-Rao bound.
\begin{figure}[t]
\centering
\includegraphics[width=85mm]{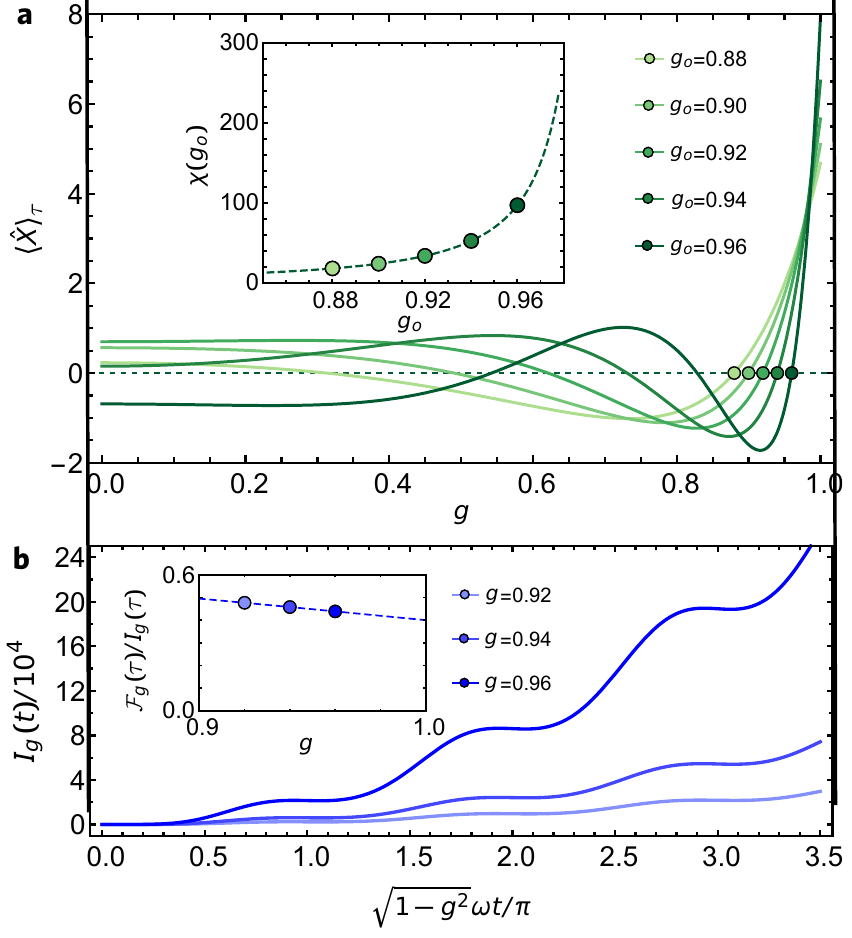}
\caption{{\bf Sensing with QRM by homodyne detection of the bosonic field.} {\bf (a)} Quadratures $\langle \hat{X}\rangle_\tau$ after an evolution time $\tau=\pi/[\omega (1-g_o^2)^{1/2}]$ as a function of $g$ . The working point $g=g_o$ are marked by filled circles. The inset shows the corresponding susceptibility at the working point $\chi(g_o)\equiv \chi_g(\tau)\vert_{g=g_o}=4\sqrt{2}\pi g_o \Delta_{g_o}^{-3/2}$. {\bf (b)} The QFI $I_{g}(t)$ as a function of the evolution time $t$. Inset: the local maximum of the inverted variance $\mathcal{F}_{g}(\tau)$ after an evolution time $\tau=2\pi/(\sqrt{\Delta_g}\omega)$ reaches the same order of $I_{g}(\tau)$.}\label{fig:quadrature}
\end{figure}
{\it Measurement protocols for QRM based sensing.---} The first method relies on quadrature measurements of the bosonic field by standard homodyne detection. Without loss of generality, we initialize the system in a product state $|\Psi\rangle=|{\downarrow}\rangle_q \otimes |\varphi\rangle_b$ with the state of the bosonic field mode $|\varphi\rangle_b=(|0\rangle+i|1\rangle)/\sqrt{2}$. After an evolution for time $t$ as governed by the Hamiltonian in Eq.\eqref{eq:H-QRM}, we perform standard measurements of the quadrature $\hat{X}=(a+a^\dagger)/\sqrt{2}$. Its mean value and variance are found to be \cite{supplement}
%\begin{equation}
  %\label{eq:X-average}
  %\begin{aligned}
   % \langle \hat{X}\rangle_t &=\sqrt{\frac{2}{\Delta_g}}\sin \left(\frac{\sqrt{\Delta_g}\omega t}{2} \right)\\
   % (\Delta \hat{X})^2 &=1+\frac{2g^2-1}{\Delta_g}[1-\cos (\sqrt{\Delta_g}\omega t )]
  %\end{aligned}
%\end{equation}
\begin{equation}
  \label{eq:X-average}
  \begin{aligned}
    \langle \hat{X}\rangle_t &=\sqrt{2}\Delta_g^{-\frac{1}{2}}\sin (\sqrt{\Delta_g}\omega t/2),\\
    (\Delta \hat{X})^2 &=1+(2g^2-1)\Delta_g^{-1}[1-\cos (\sqrt{\Delta_g}\omega t )].
  \end{aligned}
\end{equation}
We first calculate the susceptibility of the observable $\langle \hat{X}\rangle_t$ with respect to the parameter $g$, i.e. $\chi_g(t)=\partial_g\langle \hat{X}\rangle_t$, which shows divergent feature close to the critical point, see Fig.\ref{fig:quadrature}(a). To quantify the precision of parameter estimation, we denote the inverted variance as $\mathcal{F}_g(t)=\chi_g^2(t)/(\Delta \hat{X})^2$. When $\mathcal{F}_g(t)=I_g(t)$, the precision reaches quantum Cram\'er-Rao bound. The local maximums of the inverted variance with the evolution time $\tau_n= 2n\pi /\sqrt{\Delta_g} \omega $ ($n\in\mathbb{Z}^+$) \cite{supplement} is 
\begin{equation}
\mathcal{F}_g(\tau_n)=32\pi^2g^2\Delta_g^{-3}n^2 \sim I_g(\tau_n),
\label{eq:local_maximum}
\end{equation}
where the QFI $I_g(\tau_n) \simeq 64\pi^2g^2\Delta_g^{-3}n^2 \text{Var}[\hat{P}^2]_{|\varphi\rangle_b}$ is obtained from Eq.\eqref{eq:QFI-QRM}. It can be seen from Fig.\ref{fig:quadrature}(b) that $\mathcal{F}_g(\tau_n)$ is comparable with the QFI. We stress that this result holds without requiring particular initial states of the bosonic field mode \cite{supplement}.  Note that Eq.\eqref{eq:local_maximum} is achieved assuming an evolution time $\tau_n\propto \Delta_g^{-1/2}$, hence larger $\mathcal{F}_g(\tau_n)$ values would require longer evolution times.

Furthermore, we obtain the inverted variance for the estimation of the frequency $\omega$ of the bosonic field mode as $\mathcal{F}_\omega(\tau_n)=\chi_\omega^2(\tau_n)/(\Delta X)^2\simeq 2g^4\Delta_g^{-2}\tau_n^2$, where $\chi_{\omega}(\tau_n)=\partial_{\omega}\langle \hat{X}\rangle_{\tau_n}$. We remark that the average bosonic excitation is bounded by $\langle N\rangle \simeq 2g^4/\Delta_g$ \cite{supplement}. Therefore, the inverted variance can be rewritten as $\mathcal{F}_\omega(\tau_n)\simeq (1/2 g^{4})\langle N\rangle^2\tau_n^2$ for an evolution time $\tau_n\propto \Delta_g^{-1/2}$, which almost saturates the Heisenberg limit with a prefactor $(1/2 g^{4})\sim \mathcal{O}(1)$. For comparison, the adiabatic sensing protocol based on the ground state achieves the Heisenberg scaling as $\epsilon^2  N ^2 \tau^2$ (where $N\simeq
\Delta_g^{-1/2}$) with a relatively small prefactor as $\epsilon \ll 1$ due to the requirement for slow adiabatic driving \cite{Rams2018,Garbe2020} and requires a longer evolution time $\sim (1/\epsilon) \Delta_g^{-1/2}$. For the estimation of the bosonic field frequency $\omega$, the Ramsey-like interferometric protocol may also achieve the Heisenberg limit while requiring the NOON state or the superposition of vacuum and an excited Fock state, i.e., $|0\rangle+|N\rangle$ \cite{Giovannetti2004,Giovannetti2011,McCormick2019}, the preparation of which may become challenging when $N$ is large.
\begin{figure}[t]
\centering
\includegraphics[width=85mm]{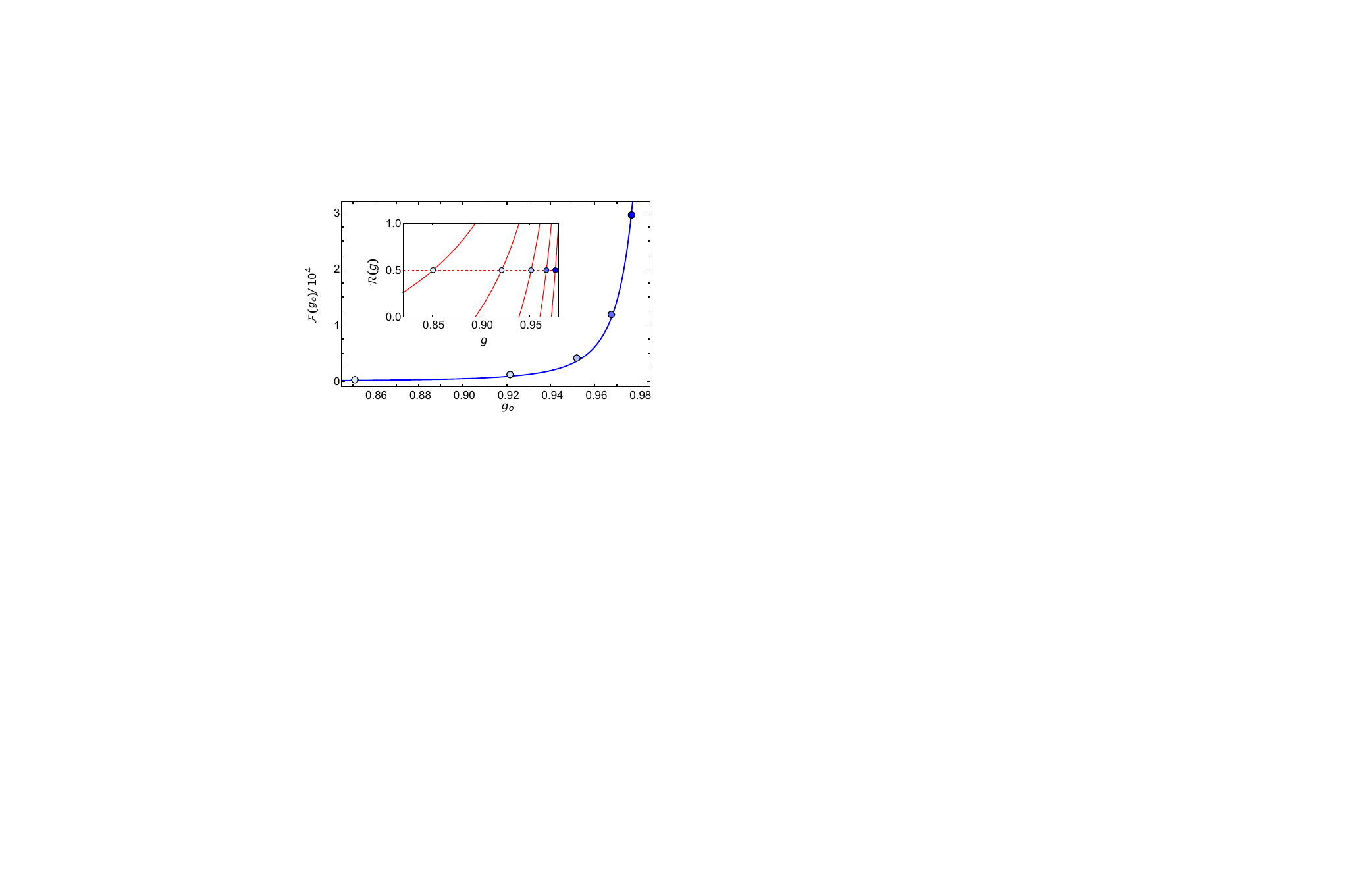}
\caption{{\bf Sensing with QRM by local measurement of the two-level system.} 
%{\bf (a)} Real (red) and imaginary parts (green) of the susceptibility $\chi^{\mathcal{G}}(g,t)$ at $g=0.97$. The second term in the susceptibility (dashed curves), see Eq.\eqref{eq:susceptibility}, makes the dominant contribution. ({\bf b}) 
The inverted variance $\mathcal{F}(g_o)$ based on the observable $\langle \sigma_x^{(q)} \rangle$ fits well with $\sim \Delta_{g_o}^{-3}$ (blue line). The inset shows the working points $g_o$ which are chosen such that $\mathcal{R}(g_o)=0.5$. The initial state is $|\Psi\rangle=(c_\uparrow|{\uparrow}\rangle_q+c_\downarrow|{\downarrow}\rangle_q)\otimes|\varphi\rangle_b$ and as an example, we choose $2c_\uparrow^* c_\downarrow =1$ and $|\varphi\rangle_b=|0\rangle$.
}\label{fig:LE}
\end{figure}
As an alternative method, one can also directly measure the two-level system to extract the parameter information \cite{Karkuszewski2002,Quan2006,Liu2019}. Without loss of generality, we start from a product initial state $|\Psi\rangle=(c_\uparrow|{\uparrow}\rangle_q+c_\downarrow|{\downarrow}\rangle_q)\otimes|\varphi\rangle_b$ with a general state of the bosonic field mode $|\varphi\rangle_b$. The observable $\langle \sigma_x^{(q)} \rangle$ of the two-level system is 
\begin{equation}
\langle \sigma_x^{(q)} \rangle =2  \mbox{Re} [ c_\uparrow^* c_\downarrow \mathcal{G}(g,t) ],
\end{equation}
where $\mathcal{G}(g,t)={}_b\langle \varphi|u_\uparrow^\dagger u_\downarrow|\varphi\rangle_b$ is the Loschmidt amplitude \cite{Karkuszewski2002}, and $u_\sigma$ is the evolution operator of the bosonic field mode when the two-level system is in the state $\vert \sigma\rangle=\vert {\uparrow}\rangle$ or $\vert {\downarrow}\rangle$. The corresponding inverted variance is $\mathcal{F}_g(\tau)=( \partial_g \langle \sigma_x^{(q)} \rangle )^2/\text{Var}[\sigma_x^{(q)}]$.
%, and $u_\sigma=\exp[-i H_{np}^{(\sigma)} t]$ with $H_{np}^{(\sigma)}=\langle \sigma|\omega [a^\dagger a+g^2(a+a^\dagger)^2\sigma_z^{(q)} /4]|\sigma\rangle_q$. 
%The susceptibility $\chi^{\mathcal{G}}(g,t)=\partial_g \mathcal{G}(g,t)\simeq \langle\varphi|u_\uparrow^\dagger %u_\downarrow h_g^{(\downarrow)}|\varphi\rangle_b \propto  \Delta_g^{-3/2}$ under the condition $\Delta_g \ll 1$, which is shown in Fig.\ref{fig:LE}(a). 
In Fig.\ref{fig:LE}, we plot the inverted variance $\mathcal{F}(g_o)=\mathcal{F}_g(\tau)|_{g=g_o}$ at the working point with an evolution time $\tau=4\pi/(\sqrt{\Delta_{g_o}}\omega)$ for the estimation of the parameter $g$ based on the observable $\langle \sigma_x^{(q)} \rangle$. Note that the working points are chosen such that $\langle\sigma_x ^{(q)} \rangle\simeq 0$ ensured by the condition $\mathcal{R}(g_o)=0.5$ \cite{supplement}, where $\mathcal{R}(g)=R(g)-\lfloor R(g)\rfloor$ with $R(g)=[(1+g^2)/(1-g^2)]^{1/2}$. It can be seen that $\mathcal{F}(g_o)$ exhibits a divergent scaling as $\Delta_{g_o}^{-3}$. It is worth to point out that the results are similar for other general initial states, such as coherent states and superposition of Fock states \cite{supplement}. We remark that the dynamic range becomes narrower when it gets closer to the critical point, which represents a general feature of criticality-based quantum sensing. Thus, the trade-off between the enhanced measurement precision and the dynamic range is a factor that one shall take into account.
{\it Experimental feasibility.---} Quantum Rabi model has been realized in a variety of quantum systems, such as cold atoms \cite{Dareau2018}, trapped ions \cite{Lv2018} and superconducting qubits \cite{Braumuller2017}. In particular, it is possible to observe quantum phase transition of QRM in the setup of a single trapped ion \cite{Puebla2017} by engineering an effective QRM Hamiltonian with the parameter $\omega /2\pi=200$ Hz, $g=2\lambda/\sqrt{\omega\omega_0}\simeq 1$ and the frequency ratio $\eta\simeq 10^3$ that can be achieved within the Lamb-Dicke regime. 

\begin{figure}[t]
\centering
\includegraphics[width=82mm]{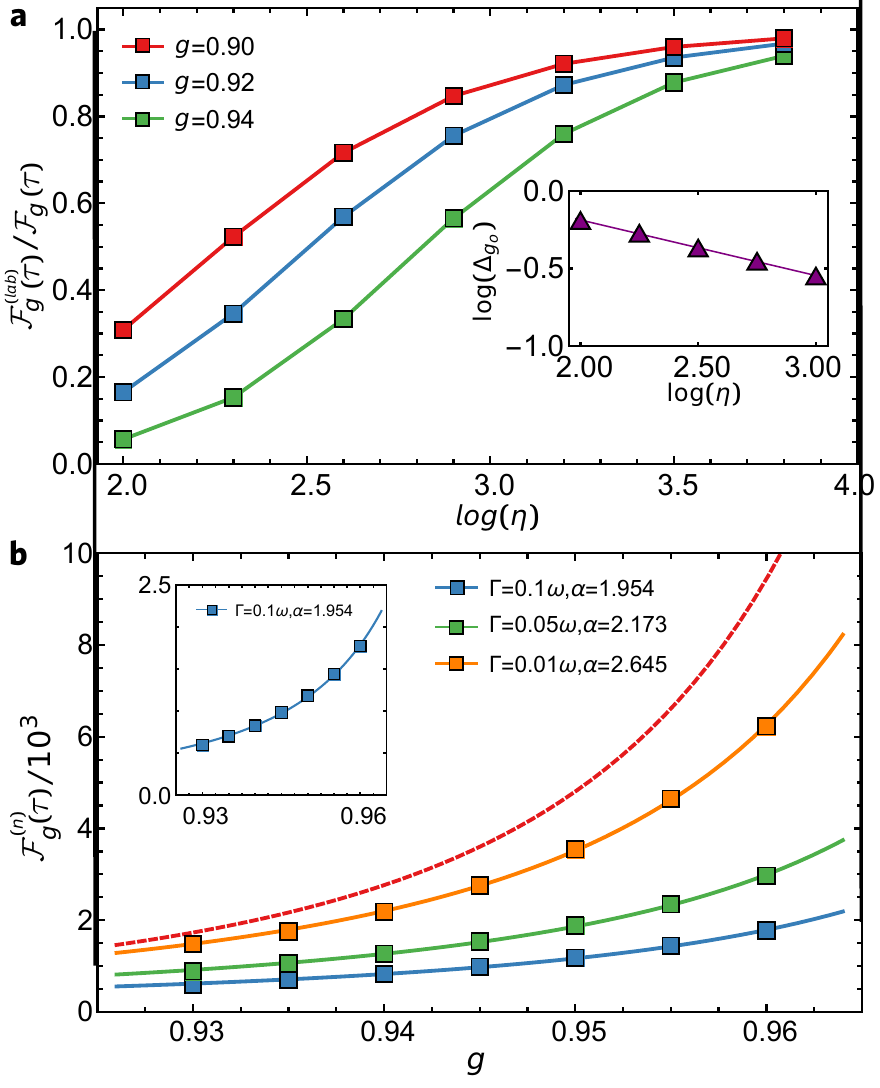}
\caption{{\bf The influence of finite frequency ratio and noise.} {\bf (a)} The ratio between the inverted variance $\mathcal{F}_g^{(\text{lab})}(\tau)$ for finite $\eta$ and the ideal value $\mathcal{F}_g(\tau)$ in the $\eta\to\infty$ limit (see Eq.\eqref{eq:local_maximum}). Inset: the parameter $\Delta_{g_o}=4(1-g_o^2)$ corresponding to the optimal working point $g_o$ at which $\mathcal{F}_g^{(\text{lab})}(\tau)$ achieves its maximum (purple triangles) is fitted by the solid curve $\log(\Delta_{g_o})=-0.358 \log(\eta)+0.528$. {\bf (b)} The inverted variance as a function of $g$ under the influence of noise with a frequency ratio $\eta=10^4$ fits well with $\mathcal{F}_{g}^{(n)}(\tau)\sim \Delta_g^{-\alpha}$. The noise parameters are chosen as $\gamma_{a,c,h}=\Gamma/2$ \cite{Puebla2017}. For comparison, the dashed curve shows the ideal result $\mathcal{F}_{g}(\tau)\sim \Delta_g^{-3}$.  The inset shows an enlargement for the case $\Gamma=0.1\omega$. In ({\bf a}-{\bf b}), the evolution time is $\tau=2\pi/(\sqrt{\Delta_g}\omega)$, and the measurement is achieved by homodyne detection of the bosonic field mode.}\label{fig:FI-lab}
 \end{figure}

Our previous analysis is presented in the limit of $\eta=\omega_0/\omega\rightarrow \infty$, which is an analog of thermodynamic limit \cite{Felicetti2020}. We address the influence of higher-order corrections in Schrieffer-Wolff expansion of the QRM Hamiltonian in Eq.\eqref{eq:H-QRM} for the experimentally accessible finite values of $\eta$. By analyzing the quadrature dynamics in more detail, we find that the leading terms of the corrections to Eq.\eqref{eq:X-average} are on the order of $\langle \hat{X}\rangle_{\text{cor}}\sim\eta^{-1}\Delta_g^{-5/2}$ and $(\Delta \hat{X}_{\text{cor}})^2\sim\eta^{-1}\Delta_g^{-3}$ \cite{supplement}. Thus, the analysis in the above section would remain valid if both $\langle \hat{X}\rangle_{\text{cor}}$ and $\Delta \hat{X}_{\text{cor}}$ are negligible as compared with $\Delta \hat{X}=1$ at the working point (i.e. $\omega\sqrt{\Delta_{g}} t =2\pi$, see Eq.\eqref{eq:X-average}). This requirement leads to a condition $\Delta_g\gg \eta^{-1/3}$ for the higher-order corrections to be negligible, which is testified by our exact numerical simulation of the inverted variance. It can be seen from Fig.\ref{fig:FI-lab}(a) that the performance of our protocol can be sustained when this condition is satisfied.  
In addition, we investigate the influence of noise which is described by the following quantum master equation as 
\begin{equation}
\begin{aligned}
&\dot{\rho}=-i[\mathcal{H}_{\text{Rabi}},\rho]+\mathcal{L}\rho,\\
&\mathcal{L}\rho =\Gamma \mathcal{D}[\sigma_z]+\gamma_{c}\mathcal{D}[\sigma_-]+\gamma_{a}\mathcal{D}[a]+\gamma_{h}\mathcal{D}[a^\dagger],
\end{aligned}
\end{equation}
where $\mathcal{D}[\hat{A}]=\hat{A}\rho \hat{A}^\dagger-\hat{A}^\dagger \hat{A}\rho/2-\rho \hat{A}^\dagger \hat{A}/2$. $\Gamma$ and $\gamma_{c}$ represent the dephasing rate and the decay rate of the two-level system, and $\gamma_{a,h} $ denote the decay rate and the heating rate of the bosonic field mode. We solve the above quantum master equation numerically and obtain the inverted variance under the influence of noise. The result is shown in Fig.\ref{fig:FI-lab}(b), in which we choose $\eta$ such that higher-order corrections are negligible in order to focus on the influence of noise. It can be seen that although the noise would decrease the achievable inverted variance, the enhancement by critical quantum dynamics still remains evident \cite{supplement} and the sub-shot-noise scaling can still be achieved with the decoherence rates (e.g. $\Gamma\lesssim 0.1\omega\simeq 2\pi \times 20$ Hz) that are experimentally reasonable, e.g. in trapped ion systems, the coherence time may exceed one second \cite{Langer2005,Niedermayr2014,Romaszko2020}. 
{\it Extension to other examples.---} The principle and analysis of quantum sensing based on critical quantum dynamics are applicable for a family of Hamiltonians exhibiting QPTs as characterized by Eq.\eqref{eq:general-condition}. Below we present two other well-known examples that can be exploited for quantum sensing enhanced by critical quantum dynamics in such a framework, namely the pumped optical parametric oscillator and the Lipkin-Meshkov-Glick model, both of which are feasible in experiments. 

The first example is the optical parametric oscillator, which interacts with the pumping field via a nonlinear optical crystal \cite{Giordmaine1965}. Under the condition that the pumping is strong and undepleted, the parametric oscillator can be described by a Hamiltonian $\mathcal{H}=\omega a^\dagger a+i\kappa[ (a^\dagger)^2-a^2]$ where $a$, $a^\dagger$ represent the bosonic annihilation and creation operators obeying the standard commutation relation $[a,a^\dagger]=1$ \cite{Walls2008}. The system satisfies the relation in Eq.\eqref{eq:general-condition} \cite{supplement}, and undergoes QPT from a real harmonic oscillator ($2\kappa/\omega<1$) to a complex harmonic oscillator ($2\kappa/\omega>1$) \cite{Fernandez2015} at the critical point $g=2\kappa/\omega=1$. In order to estimate the frequency parameter $\omega$, we can rewrite the Hamiltonian as $\mathcal{H}=\mathcal{H}_0+\omega \mathcal{H}_1$ with $\mathcal{H}_0=i\kappa [ (a^\dagger)^2-a^2]$, $\mathcal{H}_1=a^\dagger a$. The critical point $g=1$ corresponds to the parameter $\Delta_\omega=4\omega^2-16\kappa^2 =0 $. Following similar analysis, we find that the QFI scales as $\Delta_\omega^{-3}$ \cite{supplement}, which demonstrates a divergent scaling as $g\rightarrow 1$.
The second example is the Lipkin-Meshkov-Glick model \cite{Lipkin1965,Salvatori2014} with potential realizations in spin systems \cite{Mohammady2018,Capponi2019}, the Hamiltonian of which is written as 
\begin{equation}
\mathcal{H}=-\frac{1}{N}\sum_{1\leqslant i<j\leqslant N}(\sigma_i^x\sigma_j^x+\gamma\sigma_i^y\sigma_j^y)-\lambda\sum_{j=1}^N\sigma_j^z.
\end{equation}
The system undergoes QPT from paramagnetic phase ($\lambda>1$) to ferromagnetic phase ($\lambda<1$) at the critical point $\lambda=1$ for $\gamma\neq 1$. By mapping it to a large spin representation with the definitions $S_\alpha=\sum_j\sigma_j^\alpha/2$ ($\alpha=x,y,z$) and further applying the Holstein-Primakoff transformation, the LMG Hamiltonian can be approximated as $\mathcal{H}=2\lambda a^\dagger a+[\gamma (a^\dagger -a)^2-(a+a^\dagger)^2]/2$ within the regime of low excitation, i.e. $\langle a^\dagger a\rangle\ll N$ \cite{supplement}. One can verify that the LMG Hamiltonian can be written as $\mathcal{H}=\mathcal{H}_0+\lambda \mathcal{H}_1$ and also satisfies the relation in Eq.\eqref{eq:general-condition} with $\mathcal{H}_0=[\gamma (a^\dagger -a)^2-(a+a^\dagger)^2]/2$ and $\mathcal{H}_1=2 a^\dagger a$. By calculating the commutators, we obtain ${\Delta_\lambda}=16(\gamma-\lambda)(1-\lambda)$ and $\hat{D}\simeq 8(1-\gamma)(\lambda-\gamma)\hat{P}^2$, where $\hat{P}=i(a^\dagger-a)/\sqrt{2}$ \cite{supplement}. The LMG model has a vanishing energy gap $\epsilon\propto |1-\lambda|^{\nu z}$, where the critical exponents satisfy $\nu z=1/2$, belonging to the class of infinitely-coordinated models \cite{Botet1982}. We further obtain the QFI for the estimation of the parameter $\lambda$ as $I_\lambda(t)\simeq 4\Delta_\lambda^{-3} [\sin(\sqrt{\Delta_\lambda}t)-\sqrt{\Delta_\lambda}t]^2 \text{Var}[\hat{D}]_{|\Psi\rangle}$  \cite{supplement}, which becomes divergent as $\lambda\rightarrow 1$ for any initial state $|\Psi\rangle$ with a non-zero value of $\text{Var}[\hat{D}]_{|\Psi\rangle}$. 
We remark that Eq.\eqref{eq:general-condition} implies that $\Delta\sim (\lambda-\lambda_c )$. Interestingly, we note that some non-Hermitian models may also satisfy such a requirement, e.g. $H_{\lambda}=\sigma_x+i\lambda \sigma_y$, which yields $\Delta=4(1-\lambda^2)$. The two critical points at $\lambda_c \pm1$ correspond to the exceptional points. Therefore, it will be interesting to explore how to extend the present idea to quantum sensing based on this type of critical models \cite{Chen2017,Chu2020}.
{\it Conclusion and outlook.---}  To summarize, we present a dynamic framework of quantum sensing which exploits critical quantum dynamics for a family of Hamiltonians undergoing QPTs. We show that the QFI of critical quantum dynamics can exhibit a divergent scaling close to the critical point without requiring particular initial state preparation. We illustrate the principle by taking quantum Rabi model as an explicit example. The analysis and the results can be extended to more examples, such as the optical parametric oscillator and the LMG model. The present dynamic framework shows that critical quantum dynamics can provide valuable resources for engineering experimentally feasible and highly efficient criticality-enhanced quantum sensing schemes, going beyond the state-dependent divergent susceptibility. 
{\it Acknowledgements.---} We thank Prof. Martin B. Plenio and Dr. Dayou Yang for valuable comments and suggestions. This work is supported by the National Natural Science Foundation of China (No. 11874024, 11690032), the Open Project Program of Wuhan National Laboratory for Optoelectronics (No. 2019WNLOKF002).

\bibliography{reference}

 

\section*{Supplementary material}

# Supplementary material for "Dynamic framework for criticality-enhanced quantum sensing"

Yaoming Chu,[1] Shaoliang Zhang,[1, *] Baiyi Yu,[1] and Jianming Cai[1, 2, 3, †]

[1]*School of Physics, International Joint Laboratory on Quantum Sensing and Quantum Metrology,*
*Huazhong University of Science and Technology, Wuhan 430074, China*
[2]*Wuhan National Laboratory for Optoelectronics, Huazhong University of Science and Technology, Wuhan 430074, China*
[3]*State Key Laboratory of Precision Spectroscopy, East China Normal University, Shanghai, 200062, China*

## 1. Spectral feature of the Hamiltonians

In the main text, we consider a family of Hamiltonians $\hat{H}_\lambda = \hat{H}_0 + \lambda\hat{H}_1$ which satisfy the eigenvalue equation $[\hat{H}_\lambda, i\sqrt{\Delta}\hat{C} - \hat{D}] = \sqrt{\Delta}(i\sqrt{\Delta}\hat{C} - \hat{D})$ with $\hat{C} = -i[\hat{H}_0, \hat{H}_1]$ and $\hat{D} = -i[\hat{H}_1, \hat{C}]$ (see Eq.1 in the main text). Denoting the eigenstate of $\hat{H}_\lambda$ by $|\mu\rangle$ associated with an eigenvalue $\mu$, we find that

$$\hat{H}_\lambda(\pm i\sqrt{\Delta}\hat{C} - \hat{D})|\mu\rangle = (\mu \pm \sqrt{\Delta})(\pm i\sqrt{\Delta}\hat{C} - \hat{D})|\mu\rangle, \tag{S.1}$$

which implies that $|\mu \pm \sqrt{\Delta}\rangle$ is also an eigenstate of $\hat{H}_\lambda$ with the eigenvalue $\mu \pm \sqrt{\Delta}$. Iteratively, each eigenstate $|\mu\rangle$ of $\hat{H}_\lambda$ belongs to a series of eigenstates $|\mu_0 + j\sqrt{\Delta}\rangle$ ($j \in \mathbb{N}$ and $-i\sqrt{\Delta}\hat{C} - \hat{D}|\mu_0\rangle = 0$) which are equally spaced with an energy splitting $\sqrt{\Delta}$.

For example, the Hamiltonian may have an equally spaced energy spectrum, of which the energy spacing is

$$\epsilon = \sqrt{\Delta}/n \sim \sqrt{\Delta}, n \in \mathbb{Z}^+, \tag{S.2}$$

if $i\sqrt{\Delta}\hat{C} - \hat{D} = (b^\dagger)^n$. Here, the creation operator $b^\dagger$ is defined by

$$[\hat{H}_\lambda, \hat{b}^\dagger] = \sqrt{\Delta}\hat{b}^\dagger/n. \tag{S.3}$$

We note that quantum Rabi model is such an example with $n = 2$. More generally, the eigenspace of $\hat{H}_\lambda$ may consist of several manifolds, each of which contains infinite number of eigenstates that are equally spaced with an energy gap $\epsilon \sim \sqrt{\Delta}$. Therefore, we identify a family of Hamiltonians that undergo quantum phase transition, at which all energy levels collapse into a single one (or a few ones) close to the critical point $\Delta = 0$. This interesting spectral feature provides the basis for criticality-based quantum sensing without the requirement on adiabatic ground state preparation.

## 2. Derivation of the QFI of critical quantum dynamics

In the context of quantum parameter estimation, starting from an initial state $|\Psi\rangle$, the QFI for the estimation of the parameter $\lambda$ can be formulated as [1]

$$I_\lambda = 4\text{Var}[h_\lambda]_{|\Psi\rangle}, \tag{S.4}$$

where the transformed local generator $h_\lambda$ is defined as

$$h_\lambda = ie^{i\hat{H}_\lambda t}\partial_\lambda\left(e^{-i\hat{H}_\lambda t}\right) = \int_0^t e^{i\hat{H}_\lambda s}\hat{H}_1 e^{-i\hat{H}_\lambda s}\mathrm{d}s = \int_0^t \sum_{n=0}^\infty \frac{(is)^n}{n!}[\hat{H}_\lambda, \hat{H}_1]_n \mathrm{d}s = -i\sum_{n=0}^\infty \frac{(it)^{n+1}}{(n+1)!}[\hat{H}_\lambda, \hat{H}_1]_n, \tag{S.5}$$

with $[\hat{H}_\lambda, \hat{H}_1]_{n+1} = [\hat{H}_\lambda, [\hat{H}_\lambda, \hat{H}_1]_n]$ and $[\hat{H}_\lambda, \hat{H}_1]_0 = \hat{H}_1$. One can easily find that $[\hat{H}_\lambda, \hat{H}_1] = i\hat{C}$, $[\hat{H}_\lambda, i\hat{C}] = -\hat{D}$ and $[\hat{H}_\lambda, -\hat{D}] = i\Delta\hat{C}$, which results in the following commutation relations

$$[\hat{H}_\lambda, \hat{H}_1]_{2n+1} = i\Delta^n\hat{C}, \quad [\hat{H}_\lambda, \hat{H}_1]_{2n+2} = -\Delta^n\hat{D} \quad (n \in \mathbb{N}). \tag{S.6}$$

Based on Eq.(S.6), the transformed local generator can be written as (i.e. Eq.2 in the main text)

$$h_\lambda = \hat{H}_1 t - i\sum_{n=0}^\infty \frac{(it)^{2n+2}}{(2n+2)!}i\Delta^n\hat{C} - i\sum_{n=0}^\infty \frac{(it)^{2n+3}}{(2n+3)!}(-\Delta^n\hat{D}) = \hat{H}_1 t + \frac{\cos(\sqrt{\Delta}t) - 1}{\Delta}\hat{C} - \frac{\sin(\sqrt{\Delta}t) - \sqrt{\Delta}t}{\Delta\sqrt{\Delta}}\hat{D}. \tag{S.7}$$

Hence, the exact QFI is given by

$$I_\lambda = 4t^2 \text{Var}[\hat{H}_1]_{|\Psi\rangle} + g^2(t)\Delta^{-2}\text{Var}[\hat{C}]_{|\Psi\rangle} + f^2(t)\Delta^{-3}\text{Var}[\hat{D}]_{|\Psi\rangle} + I_\lambda^{(c)}, \tag{S.8}$$

where $f(t) = 2[\sin(\sqrt{\Delta}t) - \sqrt{\Delta}t]$, $g(t) = 2[\cos(\sqrt{\Delta}t) - 1]$ and the cross terms are given by

$$I_\lambda^{(c)} = 2tg(t)\Delta^{-1}\text{Cov}[\hat{H}_1, \hat{C}]_{|\Psi\rangle} - 2tf(t)\Delta^{-\frac{3}{2}}\text{Cov}[\hat{H}_1, \hat{D}]_{|\Psi\rangle} - f(t)g(t)\Delta^{-\frac{5}{2}}\text{Cov}[\hat{C}, \hat{D}]_{|\Psi\rangle}, \tag{S.9}$$

with $\text{Cov}[\hat{\Omega}_1, \hat{\Omega}_2]_{|\Psi\rangle} = \langle\Psi|\hat{\Omega}_1\hat{\Omega}_2 + \hat{\Omega}_2\hat{\Omega}_1|\Psi\rangle - 2\langle\Psi|\hat{\Omega}_1|\Psi\rangle\langle\Psi|\hat{\Omega}_2|\Psi\rangle$. Apparently, as $\lambda \to \lambda_c$ (which is defined by $\Delta = 0$), the third term in Eq.(S.8) makes the dominant contribution. It diverges as $\Delta^{-3}$ under the conditions of $\sqrt{\Delta}t \simeq \mathcal{O}(1)$ and $\text{Var}[\hat{D}]_{|\Psi\rangle} \neq 0$, thus $I_\lambda$ would exhibit a divergent scaling as (i.e. Eq.3 in the main text)

$$I_\lambda \simeq 4[\sin(\sqrt{\Delta}t) - \sqrt{\Delta}t]^2\text{Var}[\hat{D}]_{|\Psi\rangle}\Delta^{-3}. \tag{S.10}$$

More generally, we note that a divergent scaling of the QFI can in principle occur at the critical point even if the system starts from a mixed state, denoted by $\rho$. By making the spectral decomposition of $\rho$ as $\rho = \sum_{i=1}^M p_i|\Psi_i\rangle\langle\Psi_i|$, the QFI can be expressed as follows [2]

$$I_\lambda(t) = 4\sum_{i=1}^M p_i\text{Var}[h_\lambda]_{|\Psi_i\rangle} - \sum_{i\neq j}\frac{8p_ip_j}{p_i + p_j}|\langle\Psi_i|h_\lambda|\Psi_j\rangle|^2 \simeq I_\lambda^D[\sin(\sqrt{\Delta}t) - \sqrt{\Delta}t]^2\Delta^{-3}, \tag{S.11}$$

where the prefactor $I_\lambda^D$ represents the QFI of $\rho_\lambda^D = e^{-i\hat{D}\lambda}\rho e^{i\hat{D}\lambda}$ with respect to the parameter $\lambda$, namely

$$I_\lambda^D = 4\sum_{i=1}^M p_i\text{Var}[\hat{D}]_{|\Psi_i\rangle} - \sum_{i\neq j}\frac{8p_ip_j}{p_i + p_j}|\langle\Psi_i|\hat{D}|\Psi_j\rangle|^2. \tag{S.12}$$

### 3. The QFI of the QRM for general initial states

By making Schrieffer-Wolff transformation, the quantum Rabi model $\mathcal{H}_{\text{Rabi}} = \omega a^\dagger a + \omega_0\sigma_z^{(q)}/2 - \lambda(a + a^\dagger)\sigma_x^{(q)}$ in the normal phase ($g = 2\lambda/\sqrt{\omega\omega_0} < 1$) can be mapped into a block-diagonal form [3],

$$\mathcal{H} = e^{-\mathcal{S}}\mathcal{H}_{\text{Rabi}}e^{\mathcal{S}} = \frac{\omega_0}{2}\sigma_z^{(q)} + \omega\left[a^\dagger a + \frac{g^2}{4}(a + a^\dagger)^2\sigma_z^{(q)}\right] - \frac{g^4\eta^{-1}}{16}\omega(a + a^\dagger)^4\sigma_z^{(q)} + \frac{g^2\eta^{-1}}{4}\omega + \mathcal{O}\left(g^2\eta^{-2}\right), \tag{S.13}$$

where

$$\mathcal{S} = \frac{ig}{2}\eta^{-\frac{1}{2}}(a + a^\dagger)\sigma_y^{(q)} + \frac{g}{2}\eta^{-\frac{3}{2}}(a - a^\dagger)\sigma_x^{(q)} - \frac{ig^3}{2}\eta^{-\frac{3}{2}}(a + a^\dagger)\sigma_y^{(q)} + \mathcal{O}\left(\eta^{-\frac{5}{2}}\right). \tag{S.14}$$

The above Hamiltonian $\mathcal{H}$ is free of coupling terms between the spin-up and spin-down subspaces, wherein the frequency ratio $\eta = \omega_0/\omega \gg 1$. Accordingly, the initial state $|\Psi\rangle$ in the lab frame is transformed into

$$|\Psi'\rangle = e^{-\mathcal{S}}|\Psi\rangle. \tag{S.15}$$

In the interaction picture with respect to $\mathcal{H}_0 = (\omega_0/2)\sigma_z^{(q)}$, the Schrieffer-Wolff transformed Hamiltonian $\mathcal{H}$ (as $\eta \to \infty$) can be written as

$$\mathcal{H}_I = |{\uparrow}\rangle\langle{\uparrow}|_q \otimes H_{np}^{(\uparrow)} + |{\downarrow}\rangle\langle{\downarrow}|_q \otimes H_{np}^{(\downarrow)}, \tag{S.16}$$

where $H_{np}^{(\sigma)} = \langle\sigma|\omega[a^\dagger a + g^2(a + a^\dagger)^2\sigma_z/4]|\sigma\rangle_q$, $\sigma = {\uparrow}$ or ${\downarrow}$. In the spin-down subspace, we obtain that

$$H_{np}^{(\downarrow)} = \frac{\omega}{2}[\hat{P}^2 + (1 - g^2)\hat{X}^2], \tag{S.17}$$

with the quadrature operators defined as $\hat{X} = (a + a^\dagger)/\sqrt{2}$ and $\hat{P} = i(a^\dagger - a)/\sqrt{2}$. By choosing $\hat{H}_0 = (\omega/2)\hat{P}^2$ and $\hat{H}_1 = (\omega/2)\hat{X}^2$, we get

$$\hat{C} = \omega^2(\hat{X}\hat{P} + \hat{P}\hat{X})/2 \quad \text{and} \quad \hat{D} = \omega^3[\hat{P}^2 - (1 - g^2)\hat{X}^2]. \tag{S.18}$$

We define $\alpha = 1 - g^2$ and apply the formula Eq.(S.8) with $\Delta = 4\omega^2(1-g^2)$, which allows us to obtain the QFI for the measurement of the parameter $g$ as (i.e. Eq.5 in the main text)

$$I_g(t) = (\partial_g \alpha)^2 I_\alpha(t) = 4g^2 I_\alpha(t) \simeq 16g^2 \frac{[\sin(\sqrt{\Delta_g}\omega t) - \sqrt{\Delta_g}\omega t]^2}{\Delta_g^3} \mathrm{Var}[\hat{P}^2]_{|\varphi\rangle_b}, \tag{S.19}$$

where the dimensionless parameter $\Delta_g = 4(1-g^2) \ll 1$ and $|\varphi\rangle_b$ is the initial state of the bosonic field. We stress that such a scaling of $I_g(t) \propto \Delta_g^{-3}$ holds as long as $\mathrm{Var}[\hat{P}^2]_{|\varphi\rangle_b}$ is non-zero, which is valid for general initial states (see Fig.S1). This can significantly relax the requirement of initial state preparation as compared with the ground-state fidelity approach [4].

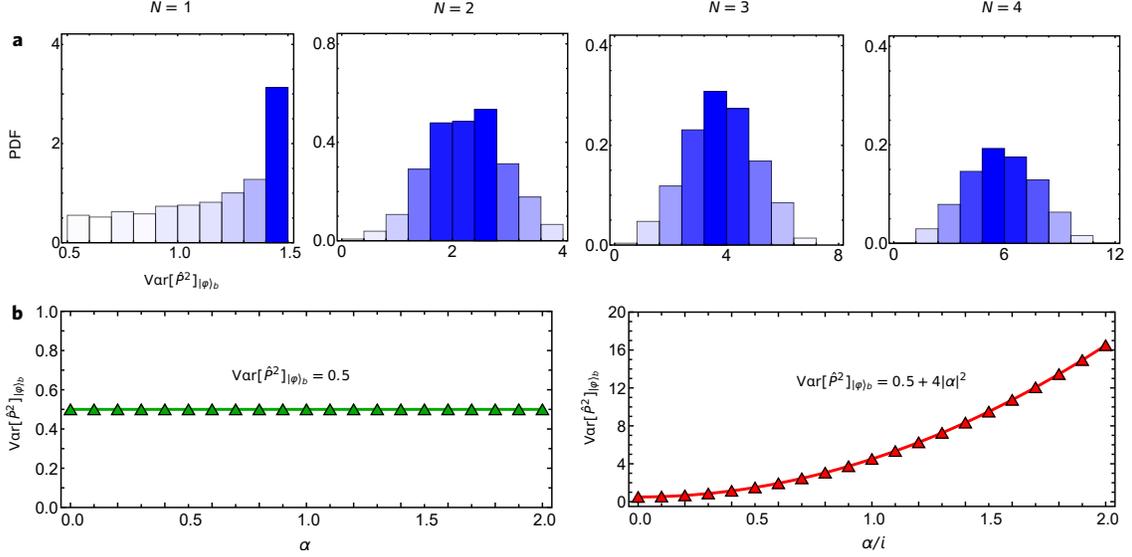

FIG. S1. **Statistics of the variance $\mathrm{Var}[\hat{P}^2]_{|\varphi\rangle_b}$ for general initial states.** (a) Probability distribution function (PDF) for $\mathrm{Var}[\hat{P}^2]_{|\varphi\rangle_b}$ (see Eq.S.19). The PDFs are derived from 5000 random initial states under the basis of $\{|n\rangle\}_{n=0}^N$ with $|n\rangle$ denoting Fock state of the bosonic field. From the left to right, we have set $N = 1, 2, 3, 4$ respectively. It can be seen that $\mathrm{Var}[\hat{P}^2]_{|\varphi\rangle_b} \gtrsim 0.5$ for almost all initial states $|\varphi\rangle_b$, which implies that the scaling of the QFI in Eq. (S.19) for the estimation of the parameter $g$ can be achieved in experiments without requiring particular initial states. (b) shows the variance $\mathrm{Var}[\hat{P}^2]_{|\varphi\rangle_b}$ of coherent states $|\varphi\rangle_b = |\alpha\rangle = \mathcal{D}(\alpha)|0\rangle$ with $\mathcal{D}(\alpha) = \exp(\alpha a^\dagger - \alpha^* a)$. In the left panel, we consider real values of $\alpha$, in which case $[\mathcal{D}(\alpha), \hat{P}] = 0$ and thus $\mathrm{Var}[\hat{P}^2]_{|\varphi\rangle_b} = 0.5$ is a constant value; while in the right panel, we choose imaginary $\alpha$ and the variance can be fitted by $\mathrm{Var}[\hat{P}^2]_{|\varphi\rangle_b} = 0.5 + 4\alpha^2$ (red curve). We remark that in both cases $\mathrm{Var}[\hat{P}^2]_{|\varphi\rangle_b} \gtrsim 0.5$ for general coherent states $|\varphi\rangle_b = |\alpha\rangle$.

## 4. Calculation details of quadrature dynamics of the QRM

In the Heisenberg picture, the equations of motion for the quadratures $\hat{X}$ and $\hat{P}$, which are governed by the effective low-energy Hamiltonian in Eq.(S.17), can be obtained as follows

$$d\hat{X}/dt = \omega\hat{P} \quad \text{and} \quad d\hat{P}/dt = -(1-g^2)\omega\hat{X}. \tag{S.20}$$

In the main text, as an example, we assume that the bosonic field is initialized in the state $|\varphi\rangle_b = (|0\rangle + i|1\rangle)/\sqrt{2}$. The dynamics of the quadrature $\hat{X}$ is given by

$$\langle \hat{X} \rangle_t = \sqrt{2}\Delta_g^{-1/2} \sin(\sqrt{\Delta_g}\omega t/2), \tag{S.21}$$

from which one can obtain the susceptibility with respect to the parameter $g$ as

$$\chi_g(t) = \partial_g \langle \hat{X} \rangle_t = 4\sqrt{2}g\Delta_g^{-3/2} \sin(\sqrt{\Delta_g}\omega t/2) - 2\sqrt{2}g\omega t\Delta_g^{-1} \cos(\sqrt{\Delta_g}\omega t/2). \tag{S.22}$$

Similarly, we get the susceptibility with respect to the field frequency $\omega$ as

$$\chi_\omega(t) = \partial_\omega \langle \hat{X} \rangle_t = \chi_g \partial_\omega g + t \cos(\sqrt{\Delta_g}\omega t/2)/\sqrt{2} = -\frac{g}{2\omega}\chi_g + t \cos(\sqrt{\Delta_g}\omega t/2)/\sqrt{2}. \tag{S.23}$$

It can be seen that both $\chi_g$ and $\chi_\omega$ exhibit critical behaviors as $\Delta_g \to 0$. In particular, when $t = \tau_n = 2n\pi/(\sqrt{\Delta_g}\omega)$ $(n \in \mathbb{Z}^+)$, we have

$$\chi_g = (-1)^{n-1}(2\sqrt{2})g\omega\tau_n/\Delta_g, \tag{S.24}$$

and

$$\chi_\omega = (-1)^n(\sqrt{2}g^2/\Delta_g - 1/\sqrt{2})\tau_n \simeq (-1)^n\sqrt{2}g^2\tau_n/\Delta_g. \tag{S.25}$$

Similar analysis shows that the linear dynamic range at a fixed working point gets narrower as $g$ goes closer to the critical point. This represents a general feature for those proposals that utilize criticality and the associated divergent behaviors for sensing. Indeed, the key advantage of criticality-based quantum sensing is to detect the minuscule changes of physical quantities. It is expected to have powerful applications in weak signal detection or precise measurement of physical parameter by tuning close to the critical point with a certain bias field based on relatively accurate parameter pre-estimation.

In order to determine the measurement precision, we also calculate the variance of the quadrature $\hat{X}$. With detailed calculations, the following equations of motion are obtained,

$$\begin{aligned}
\mathrm{d}\hat{X}^2/\mathrm{d}t &= \omega(\hat{X}\hat{P} + \hat{P}\hat{X}), \\
\mathrm{d}\hat{P}^2/\mathrm{d}t &= -(1 - g^2)\omega(\hat{X}\hat{P} + \hat{P}\hat{X}), \\
\mathrm{d}(\hat{X}\hat{P} + \hat{P}\hat{X})/\mathrm{d}t &= 2\omega^2[\hat{P}^2 - (1 - g^2)\hat{X}^2],
\end{aligned} \tag{S.26}$$

which leads to a solution of $\langle\hat{X}^2\rangle_t = 1 + 2g^2\Delta_g^{-1}[1 - \cos(\sqrt{\Delta_g}\omega t)]$. Hence, the variance of the quadrature $\hat{X}$ is (i.e. Eq.6 in the main text)

$$(\Delta\hat{X})^2 = \langle\hat{X}^2\rangle_t - \langle\hat{X}\rangle_t^2 = 1 + (2g^2 - 1)\Delta_g^{-1}[1 - \cos(\sqrt{\Delta_g}\omega t)]. \tag{S.27}$$

We note that the oscillation factor $1 - \cos(\sqrt{\Delta_g}\omega t)$ in the second term of $(\Delta\hat{X})^2$ is out phase of the one $\cos(\sqrt{\Delta_g}\omega t/2)$ in $\chi_g(t)$ (see Eq.S.22). This allows us to achieve an enhanced susceptibility while retaining a small fluctuation of the quadrature (e.g. $t = \tau_n = 2n\pi/(\sqrt{\Delta_g}\omega)$, $\Delta\hat{X} = 1$, $\chi_g \sim \Delta_g^{-1}\tau_n$). Therefore, the measurement precision of the parameter $g$ can be significantly improved. Quantitatively, the corresponding inverted variance is defined as $\mathcal{F}_g(t) = \chi_g^2/(\Delta\hat{X})^2$, which achieves its local maximums $\mathcal{F}_g(\tau_n) = 8g^2\omega^2\Delta_g^{-2}\tau_n^2 = 32\pi^2g^2\Delta_g^{-3}n^2$ (i.e. Eq.7 in the main text) at $t = \tau_n = 2n\pi/(\sqrt{\Delta_g}\omega)$ $(n \in \mathbb{Z}^+)$, see Fig.S2. Similarly, we get $\mathcal{F}_\omega(\tau_n) = \chi_\omega^2/(\Delta\hat{X})^2 \simeq 2g^4\Delta_g^{-2}\tau_n^2$. Moreover, we can calculate the average number of bosonic excitation as

$$\langle a^\dagger a\rangle_t = 1/2 + g^4\Delta_g^{-1}[1 - \cos(\sqrt{\Delta_g}\omega t)], \tag{S.28}$$

which is bounded by $\langle N\rangle \simeq 2g^4/\Delta_g$. Therefore, the inverted variance can be rewritten as $\mathcal{F}_\omega(\tau_n) \simeq (1/2g^4)\langle N\rangle^2\tau_n^2$.

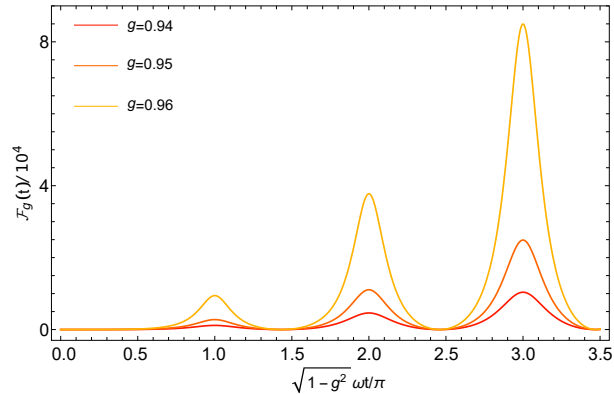

FIG. S2. **The inverted variance as a function of the evolution time**. The inverted variance $\mathcal{F}_g(t)$ shows equidistant peaks at $\sqrt{\Delta_g}\omega t/(2\pi) = \sqrt{1 - g^2}\omega t/\pi \in \mathbb{Z}^+$ with local maximums given by $\mathcal{F}_g(t) = 8g^2\omega^2\Delta_g^{-2}t^2$ (i.e. Eq.7 in the main text). We remark that the peak of the inverted variance has a certain width, thus the inverted variance can sustain a large value to a great extent for a small deviation of the evolution time.

In the main text, we analytically calculate the inverted variance $\mathcal{F}_g(t)$ achieved by homodyne detection of the bosonic field $\langle \hat{X} \rangle$ with the initial state of the bosonic field mode $|\varphi\rangle_b = (|0\rangle + i|1\rangle)/\sqrt{2}$, and show that the inverted variance $\mathcal{F}_g(\tau)$ is comparable with the QFI $I_g(\tau)$, namely $\mathcal{F}_g(\tau)/I_g(\tau) \simeq 0.4$ with the evolution time $\tau = 2\pi/(\sqrt{\Delta_g}\omega)$ (see Eq.7 in the main text). We numerically calculate the ratio $\mathcal{F}_g(\tau)/I_g(\tau)$ from 1000 random superposition states of $\{|0\rangle, |1\rangle\}$. The result suggests that the average of the ratio is $\langle \mathcal{F}_g(\tau)/I_g(\tau) \rangle \simeq 0.2$, and the inverted variance $\mathcal{F}_g(\tau)$ has a high probability to achieve values that are comparable with the QFI $I_g(\tau)$.

### 5. Derivation of local observable of the two-level system of the QRM

In the main text, we point out that one can also measure the local observable of the two-level system (e.g. $\langle \sigma_x^{(q)} \rangle = 2\text{Re}[c_\uparrow^* c_\downarrow \mathcal{G}(g, t)]$) to extract the information on the parameter. Without loss of generality, we assume $2c_\uparrow^* c_\downarrow = 1$ and thus $\langle \sigma_x^{(q)} \rangle = \text{Re}[\mathcal{G}(g, t)]$. The inverted variance for the estimation of the parameter $g$ based on the observable $\langle \sigma_x^{(q)} \rangle$ is given by

$$\mathcal{F}_g(t) = \frac{(\partial_g \langle \sigma_x^{(q)} \rangle)^2}{1 - \langle \sigma_x^{(q)} \rangle^2} = \frac{\text{Re}[\chi^{\mathcal{G}}(g, t)]^2}{1 - \text{Re}[\mathcal{G}(g, t)]^2}. \tag{S.29}$$

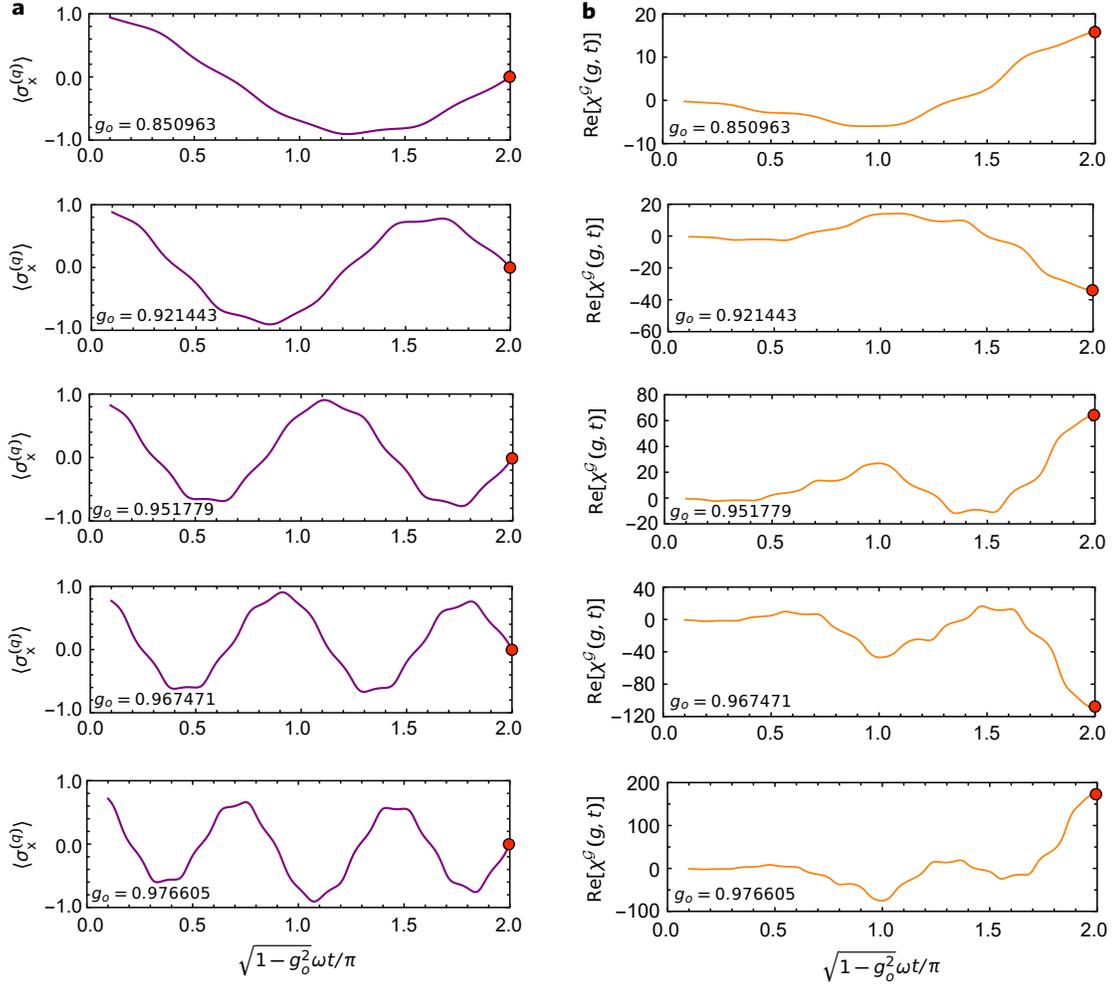

FIG. S3. **Local observable $\langle \sigma_x^{(q)} \rangle$ and its susceptibility to the parameter $g$.** (**a**) Local observable $\langle \sigma_x^{(q)} \rangle$ as a function of the evolution time at the working points $g_o$ (see Fig.2 in the main text) such that $\mathcal{R}(g_o) = 0.5$. It can be seen that $\langle \sigma_x^{(q)} \rangle \simeq 0$ after an evolution time $t = \tau \equiv 2\pi/[\omega(1 - g_o^2)^{1/2}]$ with a prominent susceptibility $\chi^{\mathcal{G}}(g, t)$. (**b**) shows the corresponding susceptibilities with respect to the parameter $g$, the absolute values of which at time $t = \tau$ increases as $g_o \to 1$. The initial state of the bosonic field mode is chosen as $|\varphi\rangle_b = |0\rangle$.

In order to choose the appropriate working point, we expand the initial state $|\varphi\rangle_b$ of the bosonic field in the eigenbasis of $H_{np}^{(\uparrow)} = \omega[a^\dagger a + g^2(a + a^\dagger)^2/4]$ and $H_{np}^{(\downarrow)} = \omega[a^\dagger a - g^2(a + a^\dagger)^2/4]$ respectively,

$$
\begin{aligned}
|\varphi\rangle_b &= \sum_m c_m^{(\uparrow)} |\phi_m^{(\uparrow)}\rangle, \\
|\varphi\rangle_b &= \sum_m c_m^{(\downarrow)} |\phi_m^{(\downarrow)}\rangle,
\end{aligned}
\tag{S.30}
$$

where $|\phi_m^{(\sigma)}\rangle = S[r_\sigma]|m\rangle$, $S[r_\sigma] = \exp[(r_\sigma/2)(a^{\dagger 2} - a^2)]$ with $r_\uparrow = -(1/4)\ln(1 + g^2)$ and $r_\downarrow = -(1/4)\ln(1 - g^2)$. As a result, we obtain

$$
\mathcal{G}(g,t) = \langle\varphi|e^{iH_{np}^{(\uparrow)}t} e^{-iH_{np}^{(\downarrow)}t}|\varphi\rangle_b = \sum_{m,n} c_m^{(\uparrow)*} c_n^{(\downarrow)} e^{i\omega t\left(m\sqrt{1+g^2} - n\sqrt{1-g^2}\right)} \langle\phi_m^{(\uparrow)}|\phi_n^{(\downarrow)}\rangle,
\tag{S.31}
$$

In particular, after an evolution time $\tau = 4\pi/(\omega\Delta_g^{1/2}) = 2\pi/[\omega(1-g^2)^{1/2}]$, we find that

$$
e^{-iH_{np}^{(\downarrow)}\tau}|\varphi\rangle_b = \sum_n c_n^{(\downarrow)} e^{-in\omega\tau\sqrt{1-g^2}}|\phi_n^{(\downarrow)}\rangle = \sum_n c_n^{(\downarrow)} e^{-i2\pi n}|\phi_n^{(\downarrow)}\rangle = |\varphi\rangle_b,
\tag{S.32}
$$

thus the Loschmidt amplitude can be rewritten as

$$
\mathcal{G}(g,\tau) = \langle\varphi|e^{iH_{np}^{(\uparrow)}\tau}|\varphi\rangle_b = \sum_{m,n} c_m^{(\uparrow)*} c_n^{(\uparrow)} e^{im\omega\sqrt{1+g^2}} \langle\phi_m^{(\uparrow)}|\phi_n^{(\uparrow)}\rangle = \sum_m |c_m^{(\uparrow)}|^2 e^{i2m\pi R(g)} = \sum_m |c_m^{(\uparrow)}|^2 e^{i2m\pi\mathcal{R}(g)},
\tag{S.33}
$$

where $R(g) = [(1 + g^2)/(1 - g^2)]^{1/2}$ and $\mathcal{R}(g) = R(g) - \lfloor R(g)\rfloor$. Under the condition $\mathcal{R}(g) = 0.5$, we get $\mathcal{G}(g,\tau) = \sum_m (-1)^m |c_m^{(\uparrow)}|^2$, which is approximately zero if the coefficient $c_m^{(\uparrow)}$ varies very slowly as $m$ increases. Therefore, we set the working points as $g = g_o$ that satisfies $\mathcal{R}(g_o) = 0.5$ (see Fig.2 in the main text) and the associated evolution time as $\tau = 4\pi/(\omega\Delta_{g_o}^{1/2}) = 2\pi/[\omega(1-g_o^2)^{1/2}]$. It can be seen from Fig.S3 that $\langle\sigma_x^{(q)}\rangle \simeq 0$ and the susceptibility $\chi^{\mathcal{G}}(g,t)$ has prominent values at these working points.

The corresponding inverted variance can be approximated as

$$
\mathcal{F}(g_o) = \mathcal{F}_g(\tau)|_{g=g_o} \simeq \mathrm{Re}[\chi^{\mathcal{G}}(g_o,\tau)]^2 \simeq 64\pi^2 g_o^2 \Delta_{g_o}^{-3} \left(\mathrm{Im}\left[\langle\varphi|u_\uparrow^\dagger u_\downarrow \hat{P}^2|\varphi\rangle_b\right]\right)^2.
\tag{S.34}
$$

With similar analysis by which we obtain $\langle\sigma_x^{(q)}\rangle \simeq 0$ at $g = g_o$ and $\tau = 2\pi/[\omega(1-g_o^2)^{1/2}]$, one can find that $\langle\varphi|u_\uparrow^\dagger u_\downarrow \hat{P}^2|\varphi\rangle_b$ takes almost a constant value that is independent of $g_o$. Thus, we can conclude that $\mathcal{F}(g_o)$ scales as $\Delta_{g_o}^{-3} \sim (1-g_o^2)^{-3}$ when $g_o \to 1$. We remark that this scaling behavior, which has been demonstrated for the initial state $|\varphi\rangle_b = |0\rangle$ of the bosonic field in the main text (i.e. Fig.2), holds for general initial states, such as superposition of Fock states and coherent states, see Fig.S4.

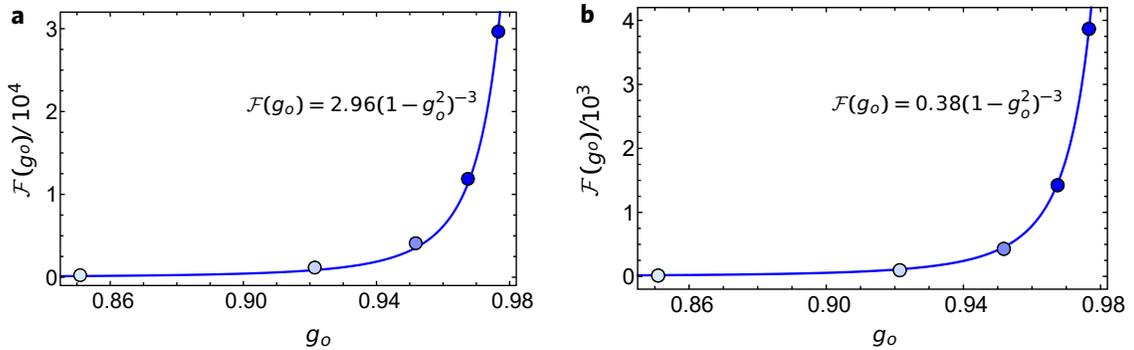

FIG. S4. **Scaling of the inverted variance for different initial states.** The working points are chosen such that $\mathcal{R}(g_o) = 0.5$ with an evolution time $\tau = 2\pi/[\omega(1-g_o^2)^{1/2}]$. As two examples, we consider a superposition of Fock states $|\varphi\rangle_b = (|0\rangle + |1\rangle)/\sqrt{2}$ (**a**) and a coherent state $|\varphi\rangle_b = |\alpha\rangle$ with $\alpha = 1$ (**b**). The inverted variance $\mathcal{F}(g_o)$ exhibits a divergent scaling as $\Delta_{g_o}^{-3} \sim (1-g_o^2)^{-3}$ for both initial states.

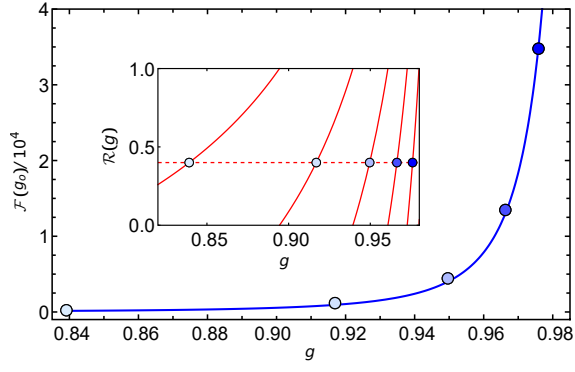

FIG. S5. **The inverted variance based on the observable** $\langle \sigma_x^{(q)} \rangle$**.** The inset shows the working points $g_o$ which are chosen such that $\mathcal{R}(g_o) = 0.4$. It can be seen that the inverted variance $\mathcal{F}(g_o)$ fits well with $\sim \Delta_{g_o}^{-3}$ (blue line). The initial state is $|\Psi\rangle = (c_\uparrow|\uparrow\rangle_q + c_\downarrow|\downarrow\rangle_q) \otimes |\varphi\rangle_b$ and as an example, we choose $2c_\uparrow^* c_\downarrow = 1$ and $|\varphi\rangle_b = |0\rangle$.

We remark that the working point $\mathcal{R}(g_o) = 0.5$ is chosen in such a way that $\langle \sigma_x^{(q)} \rangle \simeq 0$, which makes $\mathrm{Var}[\sigma_x^{(q)}] = 1 - \langle \sigma_x^{(q)} \rangle^2$ almost independent of $g$. Thus, the dependence of the inverted variance $\mathcal{F}_g(t)$ on the parameter $g$ only relies on $\partial_g \langle \sigma_x^{(q)} \rangle$ and allows us to get the scaling in a simple form as $\mathcal{F}(g_o) \sim \Delta_{g_o}^{-3}$. We would like to point out that the working point does not necessarily satisfy the condition $\mathcal{R}(g_o) = 0.5$. Similar result (i.e., the scaling $\mathcal{F}(g_o) \sim \Delta_{g_o}^{-3}$) can be obtained if we choose the working point such that $\mathcal{R}(g_o)$ takes other constant values. In Fig.S5, we show the result for the working points $g_o$ such that $\mathcal{R}(g_o) = 0.4$.

### 6. Detailed estimation of the effect of finite frequency ratio of the QRM

The normal-to-superradiant quantum phase transition of the quantum Rabi model is realized by replacing the thermodynamic limit with an infinite frequency ratio $\eta = \omega_0/\omega \to \infty$. However, only finite values of the frequency ratio $\eta \gg 1$ can be accessed in experiments, which results in corrections to Eq.(S.21) and Eq.(S.27), i.e. $\langle \hat{X} \rangle_t^{(\mathrm{lab})} = \langle \hat{X} \rangle_t + \langle \hat{X} \rangle_{\mathrm{cor}}$ and $(\Delta \hat{X}^{(\mathrm{lab})})^2 = (\Delta \hat{X})^2 + (\Delta \hat{X}_{\mathrm{cor}})^2$. Therefore, the dynamics of the quadrature in the lab frame shall be written as

$$
\begin{aligned}
\langle \hat{X} \rangle_t^{(\mathrm{lab})} &= \langle \Psi | e^{i\mathcal{H}_{\mathrm{Rabi}} t} \hat{X} e^{-i\mathcal{H}_{\mathrm{Rabi}} t} | \Psi \rangle = \langle \Psi | e^{\mathcal{S}} e^{i\mathcal{H}t} e^{-\mathcal{S}} \hat{X} e^{\mathcal{S}} e^{-i\mathcal{H}t} e^{\mathcal{S}} | \Psi \rangle \\
&= \langle \Psi | \left[ \mathbb{1} + \mathcal{O}\left(\eta^{-\frac{1}{2}}\right) \right] e^{i\mathcal{H}t} \hat{X} \left[ \mathbb{1} + \mathcal{O}\left(\eta^{-\frac{3}{2}}\right) \right] e^{-i\mathcal{H}t} \left[ \mathbb{1} - \mathcal{O}\left(\eta^{-\frac{1}{2}}\right) \right] | \Psi \rangle
\end{aligned}
\tag{S.35}
$$

with $|\Psi\rangle = |\downarrow\rangle_q \otimes |\varphi\rangle_b$. Note that the leading term $\langle \hat{X} \rangle_t = \langle \Psi | e^{i\mathcal{H}_I t} \hat{X} e^{-i\mathcal{H}_I t} | \Psi \rangle$ in the above equation exactly equals to the result in Eq.(S.21). Thus, it can be seen that the dominant contribution to the correction, resulting from the terms $e^{\mathcal{S}}|\Psi\rangle$ and $e^{-\mathcal{S}} \hat{X} e^{\mathcal{S}}$, is on the order of $\eta^{-1/2} \langle \hat{X} \rangle_t \sim (\eta \Delta_g)^{-1/2}$.

In order to evaluate the influence of the correction in the transformed Hamiltonian Eq.(S.13), i.e. $-g^4 \eta^{-1} \omega (a + a^\dagger)^4 \sigma_z / 16 = -g^4 \eta^{-1} \omega \hat{X}^4 \sigma_z / 4$, we focus on the corrected Heisenberg equations as follows

$$
\mathrm{d}\hat{X}/\mathrm{d}t = \omega \hat{P} \quad \text{and} \quad \mathrm{d}\hat{P}/\mathrm{d}t = -(1 - g^2)\omega \hat{X} + g^4 \eta^{-1} \omega \hat{X}^3,
\tag{S.36}
$$

from which we know that the last term in the second differential equation results in a correction to $\langle \hat{X} \rangle_t$ by an order of $\eta^{-1} \Delta_g^{-5/2}$. Therefore, the correction can be estimated as follows

$$
\langle \hat{X} \rangle_{\mathrm{cor}} \simeq \max \left[ (\eta \Delta_g)^{-1/2}, \eta^{-1} \Delta_g^{-5/2} \right] \simeq \eta^{-1} \Delta_g^{-5/2}.
\tag{S.37}
$$

In a similar way, we can calculate that the correction to the variance of the quadrature field $(\Delta \hat{X})^2$ is on the order of $(\Delta \hat{X}_{\mathrm{cor}})^2 \simeq \eta^{-1} \Delta_g^{-3}$. Therefore, the analysis in the limit of $\eta \to \infty$ would remain valid if the correction terms $\langle \hat{X} \rangle_{\mathrm{cor}}$ and $(\Delta \hat{X}_{\mathrm{cor}})^2$ are much smaller as compared with $\Delta \hat{X} = 1$ at the working point (e.g. $t = \tau_n$), which implies a necessary condition $\Delta_g \gg \eta^{-1/3}$.

### 7. The influence of noise on the inverted variance for more general initial states

In the main text, we investigate the influence of noise on the inverted variance based on the quadrature field $\langle \hat{X} \rangle$ starting from the initial state $|\varphi_b\rangle = (|0\rangle + i|1\rangle)/\sqrt{2}$, and find that the critical scaling of the inverted variance remains evident although the noise might reduce the enhancement by critical quantum dynamics. In order to verify that the divergent behavior of the inverted variance $\mathcal{F}_g^{(n)}(\tau)$ can sustain in the presence of noise without particular requirements on the initial states, we numerically calculate the ratio between the inverted variance under the influence of noise $\mathcal{F}_g^{(n)}(\tau)$ and the ideal inverted variance $\mathcal{F}_g(\tau)$ for more general initial states. The inverted variance $\mathcal{F}_g^{(n)}(\tau)$ in the noisy case is obtained by solving the following quantum master equation (i.e. Eq.7 in the main text)

$$\dot{\rho} = -i[\mathcal{H}_{\text{Rabi}}, \rho] + \Gamma \mathcal{D}[\sigma_z] + \gamma_c \mathcal{D}[\sigma_-] + \gamma_a \mathcal{D}[a] + \gamma_h \mathcal{D}[a^\dagger], \tag{S.38}$$

where $\mathcal{D}[\hat{A}] = \hat{A}\rho\hat{A}^\dagger - \hat{A}^\dagger\hat{A}\rho/2 - \rho\hat{A}^\dagger\hat{A}/2$. In Fig.S6, we plot the ratio $\mathcal{F}_g^{(n)}(\tau)/\mathcal{F}_g(\tau)$ for randomly chosen superposition states of $\{|0\rangle, |1\rangle\}$. The mean value of the ratio is fitted by $\mathcal{F}_g^{(n)}(\tau)/\mathcal{F}_g(\tau) = 1.44(1-g^2)^{0.3}$. We note that $\mathcal{F}_g(\tau) \sim (1-g^2)^{-3}$, which indicates a divergent scaling of the inverted variance even under the influence of noise as $\mathcal{F}_g^{(n)}(\tau) \sim (1-g^2)^{-2.7}$ without requiring particular initial states.

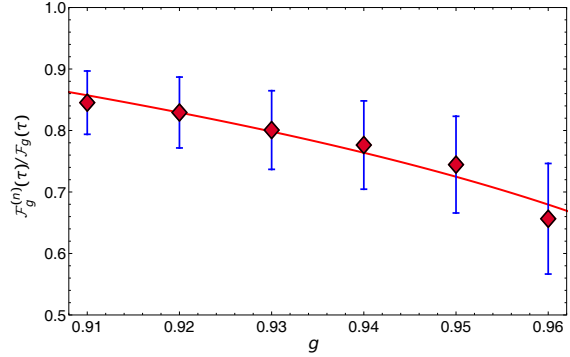

FIG. S6. **Ratio of the inverted variance $\mathcal{F}_g^{(n)}(\tau)/\mathcal{F}_g(\tau)$ for general initial states under the influence of noise.** Each point (including the mean value and standard deviation) is calculated by solving the quantum master equation starting from 200 random initial states $|\downarrow\rangle_q \otimes |\varphi_b\rangle$, with $|\varphi_b\rangle$ the superposition state of $\{|0\rangle, |1\rangle\}$. The mean value of the ratio is fitted by $\mathcal{F}_g^{(n)}(\tau)/\mathcal{F}_g(\tau) = 1.44(1-g^2)^{0.3}$ (red solid line), which indicates a divergent scaling of the inverted variance as $\mathcal{F}_g^{(n)}(\tau) \sim (1-g^2)^{-2.7}$. Here, we choose the frequency ratio $\eta = 10^4$, the evolution time $\tau = 2\pi/(\omega\sqrt{\Delta_g})$ and the decoherence rate $\Gamma = 2\gamma_{a,c,h} = 0.01\omega$.

In Fig.S7, we show the dependence of the inverted variance on the noise at a fixed working point. The results suggest that for the experimentally reasonable noise parameters $\gamma_{a,c,h}, \Gamma \lesssim 0.1\omega$ [6–8], the enhancement of the inverted variance by critical quantum dynamics remains evident $\mathcal{F}_g^{(n)}(\tau)/\mathcal{F}_g(\tau) \gtrsim 0.2$.

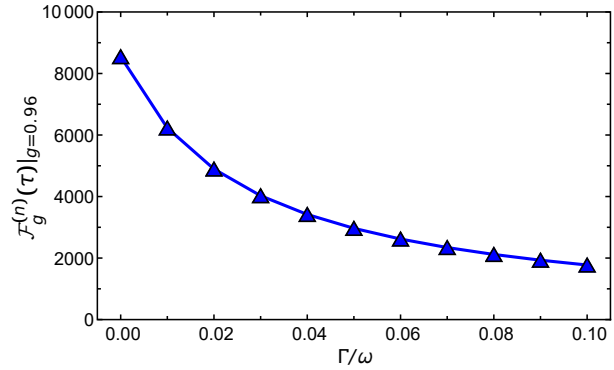

FIG. S7. **Dependence of the inverted variance on the noise at a fixed working point.** In the numerical simulation, we choose $\gamma_{a,c,h} = \Gamma/2$ [5]. It can be seen that the ratio between the inverted variance under the influence of noise and the ideal inverted variance satisfies $\mathcal{F}_g^{(n)}(\tau)/\mathcal{F}_g(\tau) \gtrsim 0.2$ when the noise level $\Gamma \lesssim 0.1\omega$. Here the parameters are chosen as $\eta = 10^4$, $\tau = 2\pi/(\omega\sqrt{\Delta_g})$, and $g = 0.96$.

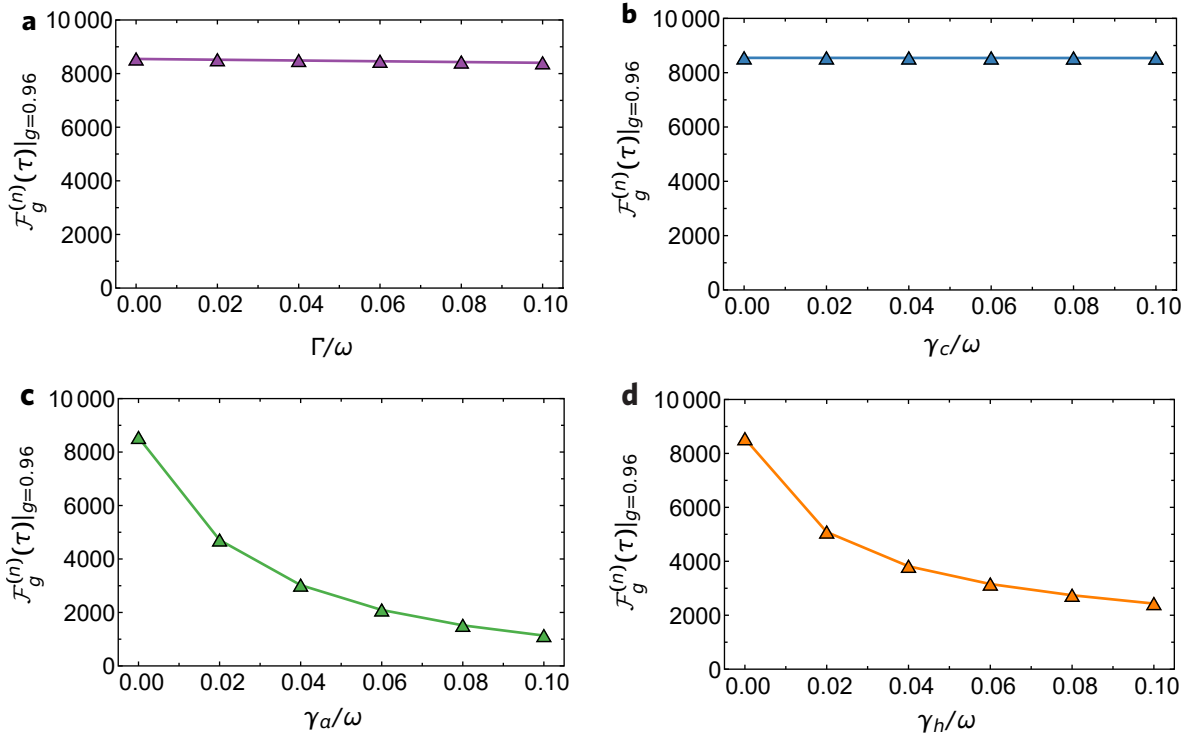

FIG. S8. **Dependence of the inverted variance on different noise processes.** It can be seen that the dephasing **(a)** and decay **(b)** of the two-level system have relatively small influences on the inverted variance $\mathcal{F}_g^{(n)}(\tau)$, which is due to the fact that the two-level system can be adiabatically eliminated. In contrast, the decay **(c)** and heating effect **(d)** of the bosonic field mode cause relatively large influence on $\mathcal{F}_g^{(n)}(\tau)$. Here the parameters are chosen as $\eta = 10^4$, $\tau = 2\pi/(\omega\sqrt{\Delta_g})$, and $g = 0.96$.

In Fig.S8, we show the dependences of the inverted variance under the influence of different noise processes at a working point, including the dephasing ($\Gamma$) and decay ($\gamma_c$) of the two-level system as well as the decay ($\gamma_a$) and heating ($\gamma_h$) of the bosonic field mode. It can be seen that the decoherence of the bosonic field mode plays a major role. Therefore, improving the coherence time of the motional states (e.g., of a single trapped ion) would be beneficial for the performance of criticality-based quantum sensing.

## 8. Calculation of the QFI for the optical parametric oscillator and the LMG model

In the example of the optical parametric oscillator, which is characterized by the Hamiltonian $\mathcal{H} = \mathcal{H}_0 + \omega\mathcal{H}_1$ with $\mathcal{H}_0 = i\kappa[(a^\dagger)^2 - a^2]$ and $\mathcal{H}_1 = a^\dagger a$, one can find that the commutators are

$$\hat{C} = -2\kappa[a^2 + (a^\dagger)^2], \quad \text{and} \quad \hat{D} = 4\omega\mathcal{H}_0 + 16\kappa^2\mathcal{H}_1 + 8\kappa^2. \tag{S.39}$$

The parameter $\Delta_\omega = 4\omega^2 - 16\kappa^2$ specifies a critical point at $\Delta_\omega = 0$ corresponding to $g = 2\kappa/\omega = 1$. Based on Eq.(S.8) (i.e. Eq.3 in the main text), the QFI for the estimation of the parameter $\omega$ is given by

$$I_\omega(t) \simeq 4[\sin(\sqrt{\Delta_\omega}t) - \sqrt{\Delta_\omega}t]^2\Delta_\omega^{-3}\text{Var}[\hat{D}]_{|\Psi\rangle}. \tag{S.40}$$

In the second example of the well-known LMG model, the Hamiltonian is written as

$$\mathcal{H} = -\frac{1}{N}\sum_{1\leqslant i<j\leqslant N}(\sigma_i^x\sigma_j^x + \gamma\sigma_i^y\sigma_j^y) - \lambda\sum_{j=1}^N \sigma_j^z, \tag{S.41}$$

where $\sigma_j^\alpha$ ($\alpha = x, y, z$) are Pauli operators at site $j$ and $\lambda$ is the effective magnetic field strength. By defining the joint spin operators $S_\alpha = \sum_j \sigma_j^\alpha/2$, this model can be mapped to a large spin that is described by the following Hamiltonian as

$$\mathcal{H} = -\frac{2}{N}(S_x^2 + \gamma S_y^2) - 2\lambda S_z. \tag{S.42}$$

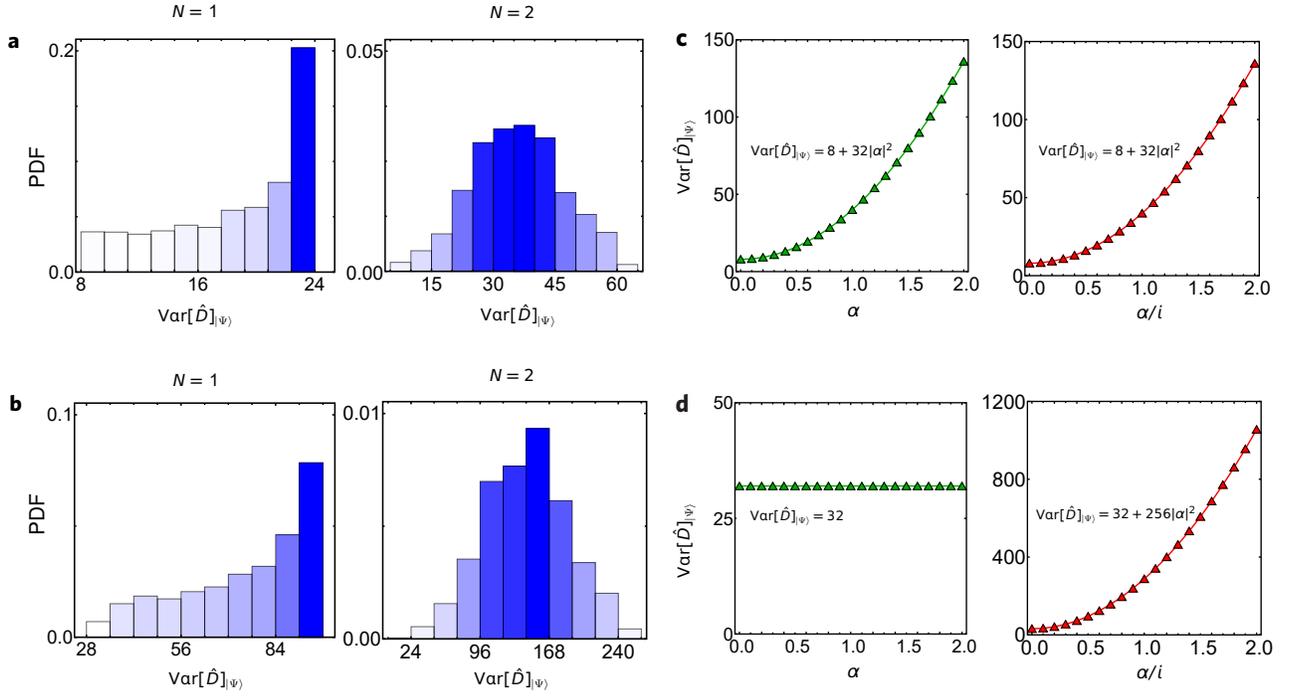

FIG. S9. **Statistics of the variance $\mathrm{Var}[\hat{D}]_{|\Psi\rangle}$ for general initial states.** (**a-b**) Probability distribution function (PDF) for $\mathrm{Var}[\hat{D}]_{|\Psi\rangle}$ are calculated from 5000 random initial states under the basis of $\{|n\rangle\}_{n=0}^{N}$ for the optical parametric oscillator (OPO) (**a**) and the LMG model (**b**). It can be seen that $\mathrm{Var}[\hat{D}]_{|\Psi\rangle} \gtrsim 8$ (OPO) and $\mathrm{Var}[\hat{D}]_{|\Psi\rangle} \gtrsim 30$ (LMG) for almost all randomly chosen initial states $|\Psi\rangle$. (**c-d**) show the result of $\mathrm{Var}[\hat{D}]_{|\Psi\rangle}$ for coherent states, i.e. $|\Psi\rangle = |\alpha\rangle = \mathcal{D}(\alpha)|0\rangle$ with $\mathcal{D}(\alpha) = \exp(\alpha a^{\dagger} - \alpha^* a)$ for real (green triangles) and imaginary (red triangles) values of $\alpha$ for the OPO model (**c**) and the LMG model (**d**). The parameters are chosen as $\omega = 1$, $g = 1$ in the OPO model and $\lambda = 1$, $\gamma = 0$ in the LMG model.

In the thermodynamic limit, the model can be exactly solved through the Holstein-Primakoff transformation and Bogoliubov transformation [9]. We focus on the Holstein-Primakoff transformation here with

$$S_+ = S_x + iS_y = \sqrt{N}\sqrt{1 - \frac{a^{\dagger}a}{N}}a, \quad S_- = S_x - iS_y = \sqrt{N}a^{\dagger}\sqrt{1 - \frac{a^{\dagger}a}{N}}, \quad S_z = \frac{N}{2} - a^{\dagger}a \tag{S.43}$$

where $a$ ($a^{\dagger}$) can be viewed as the annihilation (creation) operator for a bosonic field mode that obeys the standard commutation relation $[a, a^{\dagger}] = 1$. Within the regime of low excitation, namely $\langle a^{\dagger}a \rangle/N \ll 1$, $\sqrt{1 - a^{\dagger}a/N} \approx 1$, the LMG Hamiltonian can be approximated as

$$\mathcal{H} = 2\lambda a^{\dagger}a + [\gamma(a^{\dagger} - a)^2 - (a + a^{\dagger})^2]/2. \tag{S.44}$$

When choosing $\mathcal{H}_0 = [\gamma(a^{\dagger} - a)^2 - (a + a^{\dagger})^2]/2$ and $\mathcal{H}_1 = 2a^{\dagger}a$, we obtain the commutators (see Eq.1 in the main text) as

$$\hat{C} = 2(\gamma - 1)(\hat{X}\hat{P} + \hat{P}\hat{X}) \quad \text{and} \quad \hat{D} = 8(1 - \gamma)(\lambda - \gamma)\hat{P}^2 + 8(1 - \gamma)(1 - \lambda)\hat{X}^2, \tag{S.45}$$

where $\hat{X} = (a + a^{\dagger})/\sqrt{2}$ and $\hat{P} = i(a^{\dagger} - a)/\sqrt{2}$. Under the condition of $\gamma \neq 1$, one can see that as $\lambda \to 1$, the parameter $\Delta_\lambda = 16(1 - \lambda)(\gamma - \lambda) \to 0$ and $\hat{D} \simeq 8(1 - \gamma)(\lambda - \gamma)\hat{P}^2$. This implies that the LMG model undergoes a quantum phase transition at $\lambda_c = 1$ from paramagnetic phase ($\lambda > 1$) to ferromagnetic phase ($\lambda < 1$) for $\gamma \neq 1$. Based on Eq.(S.8) (i.e. Eq.3 in the main text), we can get the QFI as

$$I_\lambda \simeq 4\Delta_\lambda^{-3}[\sin(\sqrt{\Delta_\lambda}t) - \sqrt{\Delta_\lambda}t]^2 \mathrm{Var}[\hat{D}]_{|\Psi\rangle}. \tag{S.46}$$

We remark that the divergent scaling $I_\omega \sim \Delta_\omega^{-3}$ and $I_\lambda \sim \Delta_\lambda^{-3}$ of the QFI in both examples hold as long as the factor $\mathrm{Var}[\hat{D}]_{|\Psi\rangle} \simeq \mathcal{O}(1)$ (see Eq.S.40 and Eq.S.46), which is valid for very general initial states, see Fig.S9.

─────────

* shaoliang@hust.edu.cn
† jianmingcai@hust.edu.cn



\end{document}